\documentclass[pre,showpacs]{revtex4}
\usepackage{graphicx,graphics,color,epsfig}
\usepackage{bm}
\usepackage{amsmath}
\usepackage{amssymb}

\begin{document}

\title{Spontaneous Emergence of Modularity in a Model of Evolving Individuals and in Real Networks}

\author{Jiankui He, Jun Sun, and Michael W. Deem}

\affiliation{
Department of Physics \& Astronomy,
Rice University, Houston,Texas 77005, USA
}

\begin{abstract}
We investigate the selective forces that promote
the emergence of modularity in nature.
We demonstrate the spontaneous emergence of modularity in
a population of individuals that evolve in a
changing environment. We show that the level of modularity correlates with
the rapidity and severity of environmental change.
The modularity arises as a synergistic response to the noise in the
environment in the presence of horizontal gene transfer. 
We suggest that the hierarchical structure observed in the natural world may be
a broken symmetry state, which generically results from evolution
in a changing environment. To support our results, 
we analyze experimental protein interaction data
and show that protein interaction networks became increasingly
modular as evolution proceeded over the last four billion years.
We also discuss a method to determine the divergence time of a protein.
\end{abstract}

\pacs{87.10.-e, 87.15.A-, 87.23.Kg, 87.23.Cc}

\maketitle
 
\section{introduction}

Modularity abounds in biology.  Elements of hierarchy---modules---are
found in developmental biology, evolutionary biology, and 
ecology \cite{Shapiro2004, Shapiro2005, Lenski}.  Modularity 
is observed at levels that span
molecules, cells, tissues, organs, organisms, and societies.
At the genomic level, there are introns, exons, chromosomes, and genes.  
Moreover, there are mechanisms
to rearrange and transmit the information that is modularly
encoded at the genomic level, such as gene duplication, 
transposition, and
horizontal gene transfer
\cite{Shapiro,Goldenfeld2007}.
We define a module to be a component that can operate
relatively independently of the rest of the system. From a
structural perspective, existence of
modularity means there are more intra-module connections than
inter-module connections.  From a functional perspective, a
module is a unit that can perform largely the same function in
different contexts.
Modularity has been characterized in a variety of network
systems by physical methods \cite{Lipson2002, Wolde}. 
Selection for stability, for example, has been shown to select
for modular networks \cite{Lipson2004}.
A dictionary of constituent parts, or
network motifs, has been identified for 
the transcriptional regulation network
of \emph{E.\ coli} \cite{Alon2002}.
And once modularity has arisen, so that the goals a species
face become modular, modularly varying goals have been shown
to select for modular structure \cite{Alon2005}.
Horizontal gene transfer has been suggested to be essential 
to the evolution of a universal genetic code \cite{Goldenfeld2006}.

How does modularity arise in nature?
It has been suggested that by being modular,  a system
will tend to be both more robust to perturbations and more
evolvable \cite{Doyle, Kitano, Cluzel}.    
It has further been suggested that there is a selective
pressure for positive evolvability in a population of
individuals in a changing environment \cite{Earl}. 
Thus, we have hypothesized that modularity arises spontaneously
from the generic requirement that a population of individuals
in a changing environment be evolvable \cite{Deem2007}.
Support for this hypothesis had been elusive
\cite{Gardner2003}.

In this article, we extend the analysis presented in \cite{jun}, 
as well as discuss experimental data.
In Section II,  we introduce the spin glass model 
for the replication rate in evolution.
In Section III, we show spontaneous evolution
of hierarchy in a system under changing environmental conditions
with horizontal gene transfer.  Specifically, we show that in the presence of
horizontal gene transfer, environmental change leads to the
spontaneous emergence of modularity in a generic model of
a population of evolving individuals. 
The model describes evolution in a rugged landscape, when the
environment is changing and when horizontal gene transfer is possible.
Modularity grows spontaneously even when the horizontal gene transfer
event is of a random length and starting location.
In Section IV, we discuss
experimental evidence
in support of our simulation results. 
First we review the evidence 
showing that the bacterial metabolic networks in more variable 
environments are more modular. Next, we show using a measure of 
protein divergence time
that modularity in protein interaction networks and protein domain 
interaction networks appears to have increased with time. 
We conclude in Section V.  Additional details are presented
in Appendices.

\section{spin glass model of evolution}
  To represent the 
replication rate, or
microscopic fitness, of the individuals, we use a spin glass model
that has proved useful in previous studies of evolution
\cite{Kauffman,Deem,Sun}.   
The choice of a spin glass model, with many local fitness optima, is motivated
by our assumption that evolution occurs on a rugged landscape.  In other
words, our results pertain
only to those evolutionary processes that occur on
such rugged fitness landscapes.  A spin glass model
generically represents such rugged fitness landscapes.  We present
illustrative results for some numerical values
of the parameters in the model.  The qualitative
nature of our results are insensitive to the specific values of
these parameters.
In this model, spontaneous emergence of modularity, 
however, generically occurs for a population of evolving individuals and
depends only on the presence of a changing environment and the presence 
of horizontal gene transfer.
This spin glass model is appropriate because
it provides a rugged, difficult landscape upon which evolution struggles
to occur, and so there can be a pressure for more efficient
evolutionary structures to arise.  This rugged landscape of this
model is expected to
reproduce the slow dynamics of 
evolution \cite{Anderson,Stein,Kauffman,Perelson,Mezard},
and we have used correlated random energy models in a number of
protein evolution \cite{Bogarad,Earl} and immune system evolution
studies \cite{Deem,Sun,Sun2006}.
  There are three time scales
in our system: the fastest time scale of sequence evolution
of population as descendants replace parents, the intermediate time scale
of environmental change, and the longest time scale of 
the change to the structure of interactions between elements of
the sequence space.
The symmetry of a uniformly random structure
is broken by the spontaneous emergence of modular structure as a response
to environmental change.

We use the following spin glass form
for the microscopic fitness of proteins
in our system (for a discussion on the spin glass
approach to evolution, see \cite{Deem,Earl,Sun,Sun2006}):
\begin{equation}
H^\alpha(s^{\alpha,l})=
\frac{1}{2\sqrt{N_D}}\sum_{i \neq j} \sigma_{i,j}(s^{\alpha,l}_i, s^{\alpha,l}_j) \cdot
\Delta_{i,j}^\alpha,
\label{1}
\end{equation}
where $s^{\alpha,l}_i$, $1 \le i \le N$,
 is a string of length $N$ that specifies the
identity of ``individual'' $l$.
The term $s^{\alpha,l}_i$ may represent the amino acid at position $i$ within
the sequence of a protein, 
the label of a protein at gene $i$ in the genome,
or the type of transcriptional regulatory element
at non-coding position $i$.
  For these three examples, the
modularity that may develop represents the formation of secondary 
structure, protein-protein interaction motifs, or 
regulatory structure, respectively.
The different individuals are enumerated by $l$, with
$1 \le l \le N_{\rm size}$, where we have
$ N_{\rm size}$ different individuals.
The different possible forms of the structure of the interaction
between the $s^{\alpha,l}_i$ are enumerated by
$\alpha$, $1 \le \alpha \le D_{\rm size}$, where we choose $D_{\rm size}$
possible structures.
These structures of the interaction represent, for example, the 
protein fold, protein interaction pathways, or constraints on regulation.
The term 
$\sigma_{ i,j}(s_i, s_j)$, is the numerical value of the interaction matrix,
symmetric in $i$ and $j$,
whose elements are each taken from
a Gaussian distribution with zero mean
and unit variance.  It differs for each
$i$, $j$, $s_i$, and $s_j$.  The effect of the environment
is encoded by these random couplings.  When the environment
changes with severity $p$, each of the couplings is
with probability $p$ randomly redrawn from the Gaussian distribution.
The term $\Delta_{ i,j}^\alpha$ defines the structure of the
interaction, i.e.\ the contact matrix, or connections in structure, 
for structure $\alpha$.
The matrix is
symmetric, with elements 0 or 1.
In order to guarantee that the emergence of modularity
comes from redistribution of connections rather than
an increase in the number of connections, 
we constrain $\sum_{ i > j+1} \Delta_{ i,j}^\alpha =
N_D=346$.  Any value of $N_D$ such that
the connection matrix is neither all unity nor all zero
would give qualitatively similar results.
We take $\Delta_{ i,i}^\alpha=0$ and $\Delta_{
i, i \pm 1}^\alpha=1$.

Horizontal gene transfer is assumed, for specificity, to transfer any of the
12 blocks of length 10 in the sequence (i.e.\ sequence elements 1 \ldots 10, 
11 \ldots 20, 21 \ldots 30, etc.). 
This horizontal gene transfer event represents transfer of pieces of genes,
collections of genes, or stretches of non-coding regulatory information
between individuals.
Modularity is defined,  conjugate to the horizontal gene transfer event, to be
the number of connections
within the 12 $10 \times 10$ blocks along the diagonal
\begin{equation}
M^\alpha=\sum_{k=0}^{11} \sum_{ i=1, j=i+2}^{10} \Delta_{10 k+i, 10k +j}^\alpha,
\label{eq:mod}
\end{equation}
so that $i,j$ are within the $1+k^{th}$ diagonal block of size
$10$.  Even a random distribution of contacts
will have a non-zero absolute modularity, $M_0$, and so
it is the excess modularity that measures the degree of
spontaneous symmetry breaking, $\delta M^\alpha=M^\alpha-M_0$.
Emergence of modularity means that as a result of evolution,
connections in structure are not evenly distributed
between positions.
The interactions are greater in the local, diagonal blocks
than in the rest of the matrix, and so
$\delta M^\alpha > 0$. 
In other words, $\delta M^\alpha$ is
the order parameter of spontaneous symmetry breaking
of the approximately uniform distribution of contacts, and in the
broken symmetry phase, where the distribution of contacts is
not uniform, and $\delta M^\alpha \neq 0$.

In order to see the emergence of modularity, we need a set
of individuals in a changing environment.  Moreover, since
we want to watch the evolution of the structural connections 
$\Delta_{i,j}^\alpha$, we need
a population of these sets, each set with a different
$\Delta_{i,j}^\alpha$.  We take the population size to be
$D_{\rm size}=300$ different structures, $1 \le \alpha \le D_{\rm size}$,
and each given structure has  a set of
 $N_{\rm size}=1000$
different sequences, $1 \le l \le N_{\rm size}$,
 associated with them. 
In total
there are
$D_{\rm size}\times N_{\rm size}=3\cdot10^5$ different individuals,
replicating at the rate given by the microscopic fitness 
associated with its set [see Eq.\ (\ref{fitness}), below].
The average excess modularity is
given by $\delta M =  M - M_0 =
\frac{1}{D_{\rm size}}\sum_{\alpha=1}^{D_{\rm size}} M^\alpha - M_0$.

The structures, $\Delta_{i,j}^\alpha$,
are initialized by first randomly generating
one such structure with $N_D=346$ and a certain $M$. 
We then obtain the full set of $D_{\rm size}$ structures by
evolution away from this structure. 
Two elements of $\Delta_{ i,j}^\alpha$ with opposite
status are randomly chosen,
and the status of each is flipped from $1 \to 0, 0 \to 1$.
These mutations are done $n$ times, where $n$ is a Poisson
random number with mean 2.
The sequences,
 $s^{\alpha, l}_i, 1 \le i \le N$, of each individual
are initialized by random assignment.

The
evolution in our simulation involves three levels of change.
The most rapid change occurs by evolution of the sequences
through mutation and horizontal gene transfer.
 The selection on this level is based on the microscopic fitness.
For each structure $\Delta_{ i,j}^\alpha$, at each round, all the $N_{\rm
size}$ associated sequences undergo mutation,
horizontal gene transfer,
and selection.  The Poisson mutation process 
changes on average $2.4$ values of the $s^{\alpha,l}_i$ in the
sequence, which are randomly
selected and assigned a random new value. 
In horizontal gene transfer, two
randomly selected sequences from the population associated with one
structure attempt to exchange each of the 12
sequence fragments between $10k+1$ and $10k+10$
(of length $10$) with probability $0.1$.  Thus, the horizontal gene 
transfer rate
and the mutation rate are roughly equal \cite{Park}.
The qualitative behavior of the results does not depend on the exact
mutation rates.
All the sequences undergo attempted horizontal gene transfer
to make the
new population.
Pairs of sequences in the population associated with one structure
are chosen, until all sequences have been chosen.
This process is a model 
of horizontal gene transfer or recombination. 
The 50\%\ sequences with the lowest energy 
are selected and randomly duplicated to recover the population of
$N_{\rm size}$ for the next round; the
microscopic replication rate, or fitness, for sequence $\alpha,l$ in
structure $\alpha$ is
\begin{equation}
r^\alpha(s^{\alpha,l}) = 
2 \theta [H^\alpha_{{\rm N_{size}}/2} - H^\alpha(s^{\alpha,l})] \ ,
\label{fitness}
\end{equation}
 where
$\theta(x)$ is the Heavyside step function.
Mutation and
selection are repeated $T_2$ rounds.

The next most rapid change is that of the environment,
which occurs with severity $p$ and frequency $1/T_2$.
That is, the set of individuals evolve for $T_2$ rounds
in each given environment, and then the 
environment changes.
During the environmental change, the 
elements of the interaction matrix
$\sigma_{i,j}$ change with probability $p$.

The slowest level of change is the structural evolution.
The selection at this level is based on
the cumulative fitness of the set of individuals with a given structure,
averaged over $T_3 = 10^4 T_2$ environmental changes. 
That is, we sum the average energy of the sequence set of each
structure at the end of each environment for $T_3/T_2$ times and use
this cumulative fitness to determine the replication rate
of the structures, quantifying their performance in responding to
environment changes. 
The structures with the best $5\%$
cumulative fitness are selected and randomly amplified to
make the new population of $D_{\rm size}$ structures, $\Delta^\alpha_{ij}$.
The structure
population also undergoes mutation. 
As with the initial construction,
two elements of $\Delta_{ i,j}^\alpha$ with opposite
status are randomly chosen,
and the status of each is flipped from $1 \to 0, 0 \to 1$.
These mutations are done $n$ times, where $n$ is a Poisson
random number with mean 2.
The mutated
structures, $\Delta_{ i,j}^\alpha$,  are used for the next
$T_3$ rounds of evolution.

\section{spontaneous emergence of modularity}

\begin{figure}[t]
\begin{center}
\epsfig{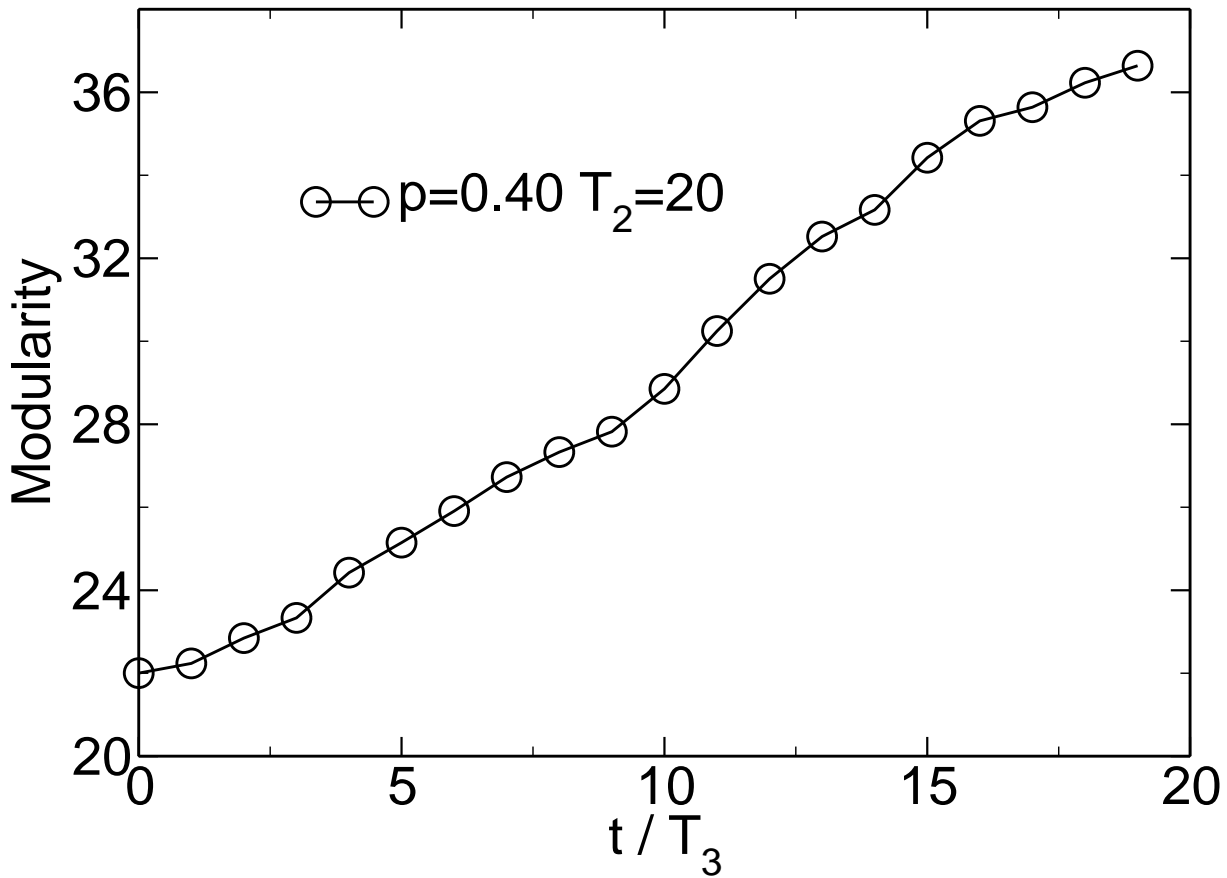}
\end{center}
\caption{Spontaneous emergence of excess modularity,
$M > M_0 = 22$, from
a state with no excess modularity, 
$M=M_0$. The random, symmetric distribution of
structural connections is spontaneously broken as the system
evolves. Here $T_2=20$, $T_3= 10^4 \times T_2$,
and the severity of environmental change is 
$p=0.40$. }
\label{fig:MandT-O-0}
\end{figure}

\begin{figure}[t]
\begin{center}
\epsfig{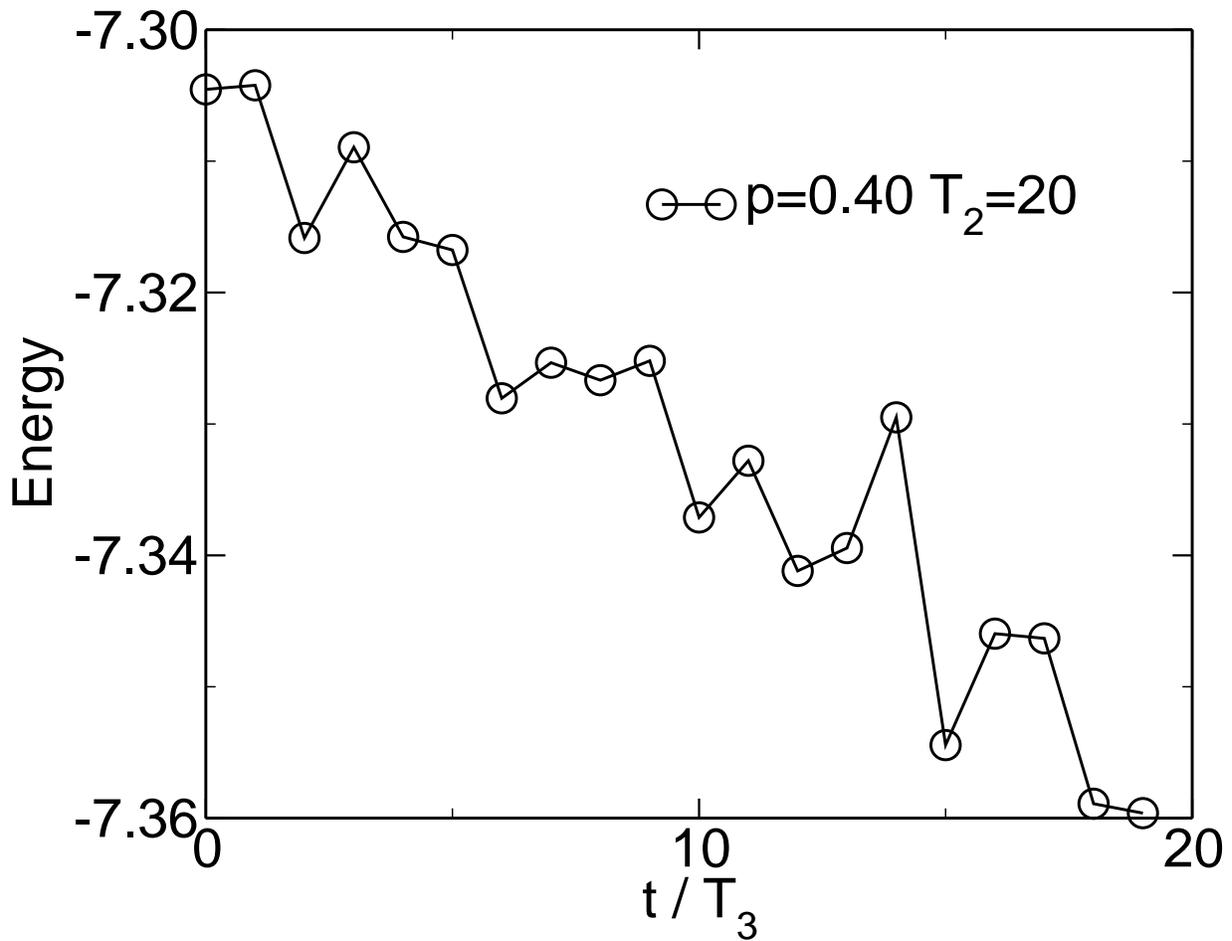}
\end{center}
\caption{Improvement in the energy as time increases and as
the modularity grows, as shown by 
Fig.~\ref{fig:MandT-O-0}. 
 Here the severity of environmental change is $p = 0.4$, and the
period of change is $T_2 = 20$.
Here, and in all figures, $T_3=10^4 \times T_2$.
\label{fig:MandT-O-2}
}
\end{figure}

\begin{figure}[t]
\begin{center}
\epsfig{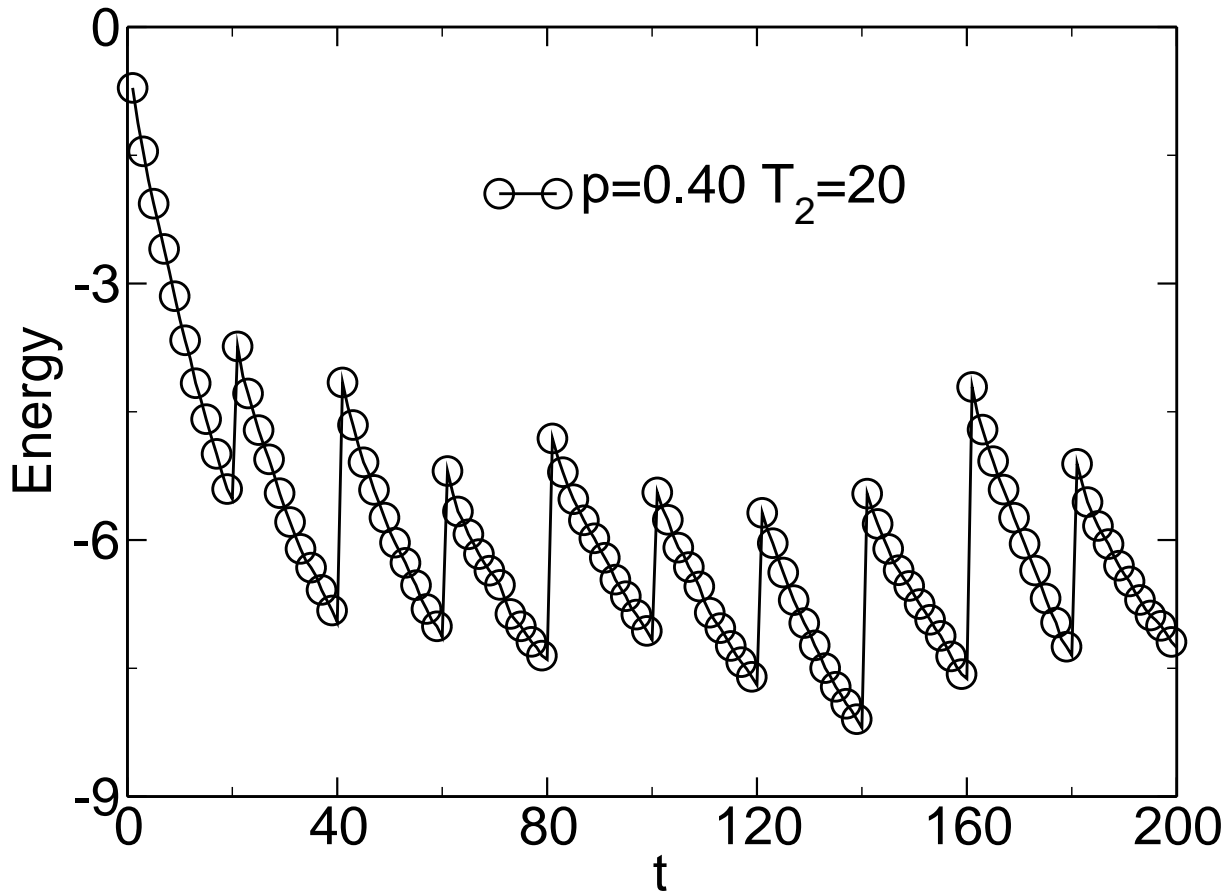}
\end{center}
\caption{Energy improvement as evolution proceeds within each environment 
and large energy disruption due to environmental changes. 
 Here the severity of environmental change is $p = 0.4$, and the
period of change is $T_2 = 20$.
}
\label{fig:MandT-O-5}
\end{figure}

\begin{figure}[t]
\begin{center}
\epsfig{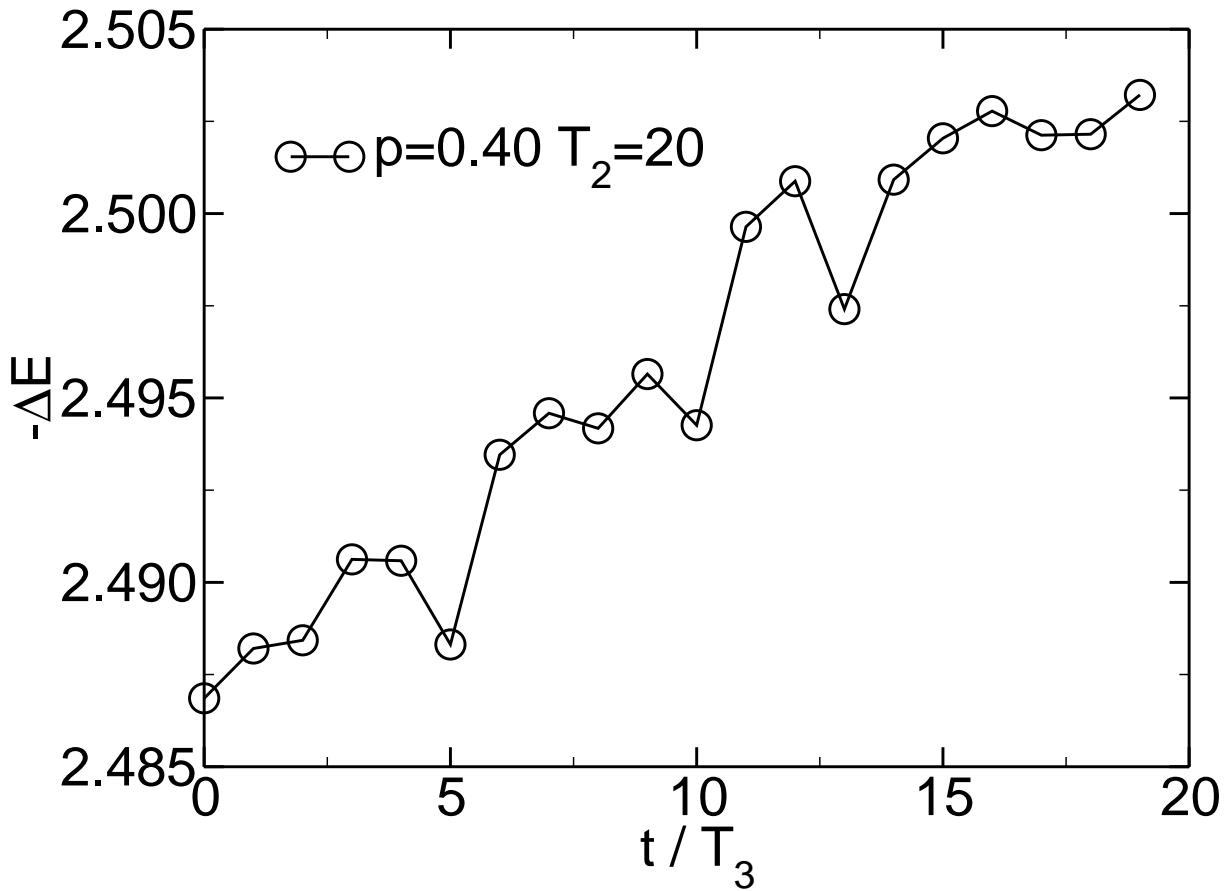}
\end{center}
\caption{Improvement of evolvability, or evolved improvement of the energy
in one environment, as the modularity grows. 
 Here the severity of environmental change is $p = 0.4$, and the
period of change is $T_2 = 20$.
\label{fig:MandT-O-3}
}
\end{figure}

\begin{figure}[t]
\begin{center}
\epsfig{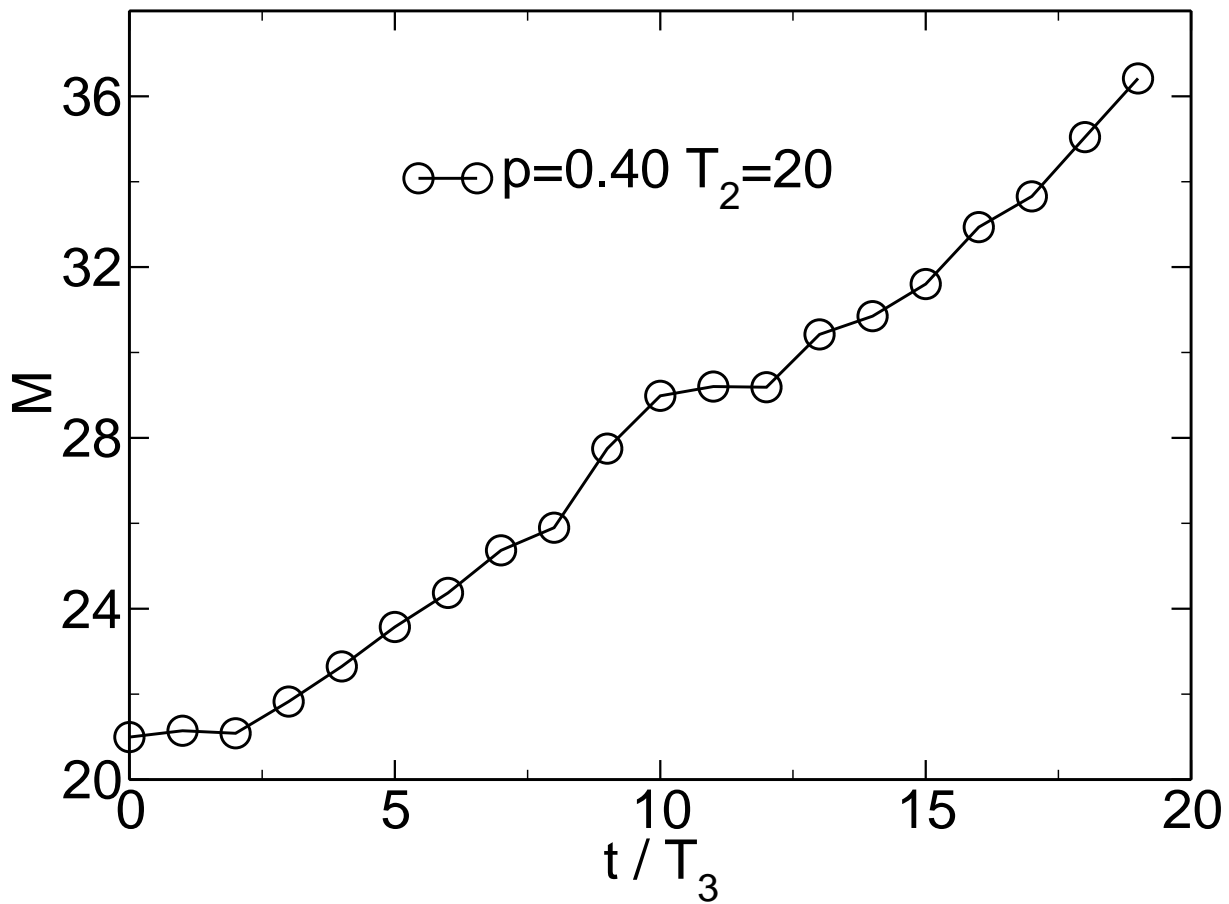}
\end{center}
\caption{Spontaneous emergence of excess modularity,
$M > M_0 = 22$, from an initial scale-free network ($\gamma = 3$)
with $M=M_0$.
 Here $T_2=20$, and the severity of environmental change is 
$p=0.40$. 
\label{fig:MandT-O-1}
}
\end{figure}

In Fig.~\ref{fig:MandT-O-0}, we show the spontaneous emergence of
modularity from the symmetric, random state 
of no excess modularity, $M = M_0 = 22$.
Since the system is initially quite far from the steady state modularity,
the growth of the excess modularity with time is roughly linear.
The excess modularity is the order parameter for this system, and its 
growth shows that the system is in a broken symmetry phase
with modular structure under these conditions. In Fig.~\ref{fig:MandT-O-2}, we
show the energy decreases
 as modularity grows, \emph{i.e.}\ the stability of the structure
is increasing.  In Fig.~\ref{fig:MandT-O-5}, we show the change of energy 
with time in more detail. Compared with Fig.~\ref{fig:MandT-O-2}, the time
scale ($x$-axis) in Fig.~\ref{fig:MandT-O-5} is much smaller
($T_3=2 \times 10^5$).
The energy decreases during each constant environment.
The environment
changes each $t=20$ steps.  Immediately after the change of environment,
the individual sequences are not as well adapted, and so the 
the energy increases sharply.
As the sequences adapt in the new environment,
the average energy of the population decreases.
In Fig.~\ref{fig:MandT-O-3}, the response function, or evolvability,
$\Delta$E, is shown as a function of time.
By evolvability,
we mean the rate of change in a new environment \cite{Earl}.
We 
observed the growth of evolvability as the modularity grows 
in Figs.~\ref{fig:MandT-O-1} and \ref{fig:MandT-O-3}.

Interestingly, the growth of modularity is identical for an initial
contact matrix that is power-law distributed.
Many
biological networks appear scale free, at least over a 
limited range of connectivity \cite{barabasi_networkbiology}, with a power-law
degree distribution. Here, we choose the Barab\'{a}si \textit{et al.}' method \cite{barabasi_networkbiology}
to generate an initial contact matrix that is power-law distributed
with $\gamma=3$. In Fig.\ \ref{fig:MandT-O-1}, we show the growth of
modularity that is nearly identical to that of Fig.\ \ref{fig:MandT-O-0}.

\begin{figure}[t]
\begin{center}
\epsfig{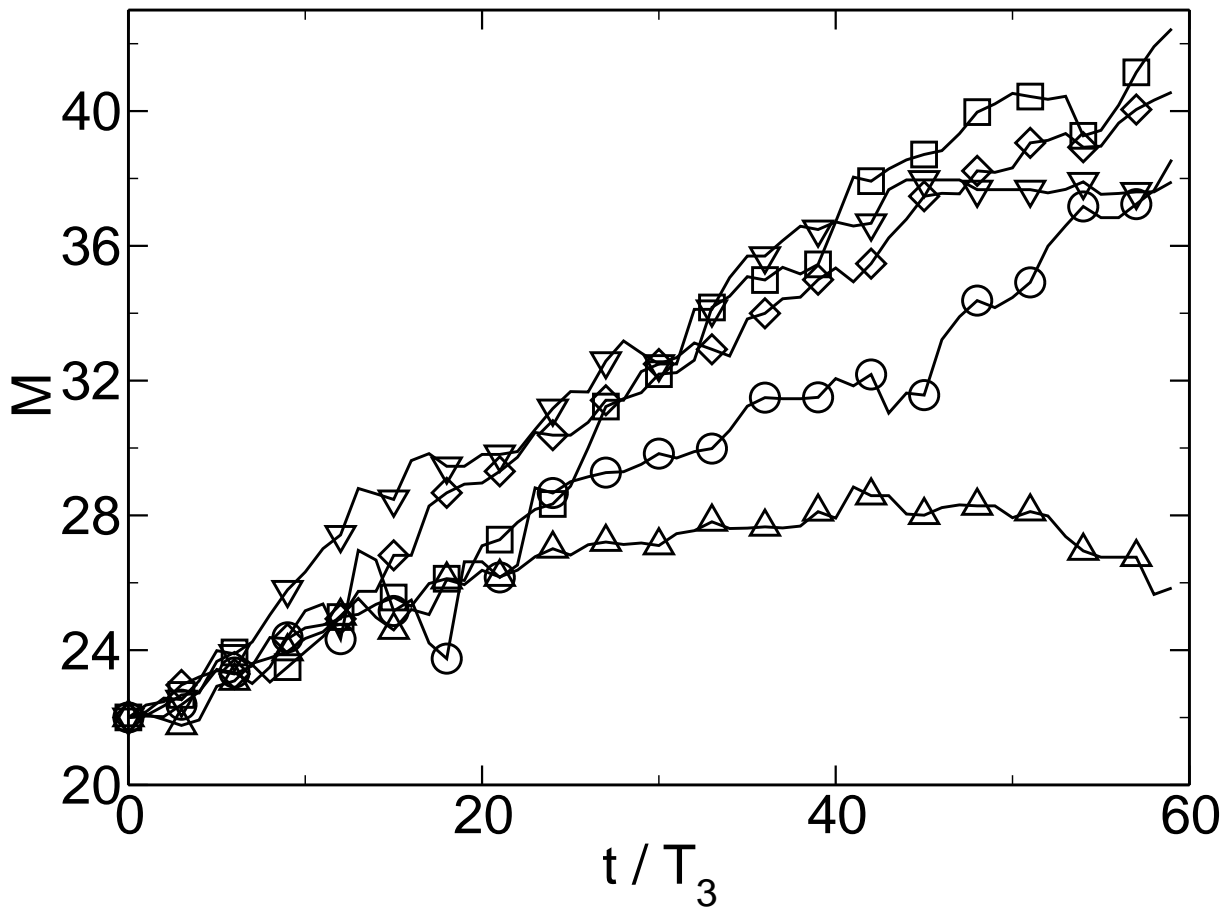}
\end{center}
\caption{Emergence of modularity as a 
result of a horizontal gene transfer operator with a Poisson random 
swap length 
and uniform random starting position. 
Shown are data for an average swap length of 
10 ($\bigcirc$),
20 ($\Box$), 
20 ($\Diamond$), 
5 ($\bigtriangleup$),
and 40 ($\bigtriangledown$)
 with 12, 6, 12, 24, and 3 attempted swaps, respectively,
of probability 0.1 per sequence pair.
Here $T_2=20$, and the severity of environmental change is 
$p=0.40$. }
\label{fig:MandT-O-4}
\end{figure}

The spontaneous emergence of modularity is a general result. 
In Fig.~\ref{fig:MandT-O-4}, we show the excess modularity still grows, even if 
the gene transfer starts at a uniformly random position and
swaps a random length of sequence. 
the original assumption of fixed length and position,
however, is biologically motivated.  
If we take the specific instance of the model to indicate formation of
secondary structures or protein-protein interactions, then 
if the blocks are exons, and
the ratio of non-coding to coding DNA is large, then typical recombination
or horizontal gene transfer will transfer an integer number of complete
exons, which is our horizontal gene transfer operator
of fixed length and position. 

\begin{figure}[t]
\begin{center}
\epsfig{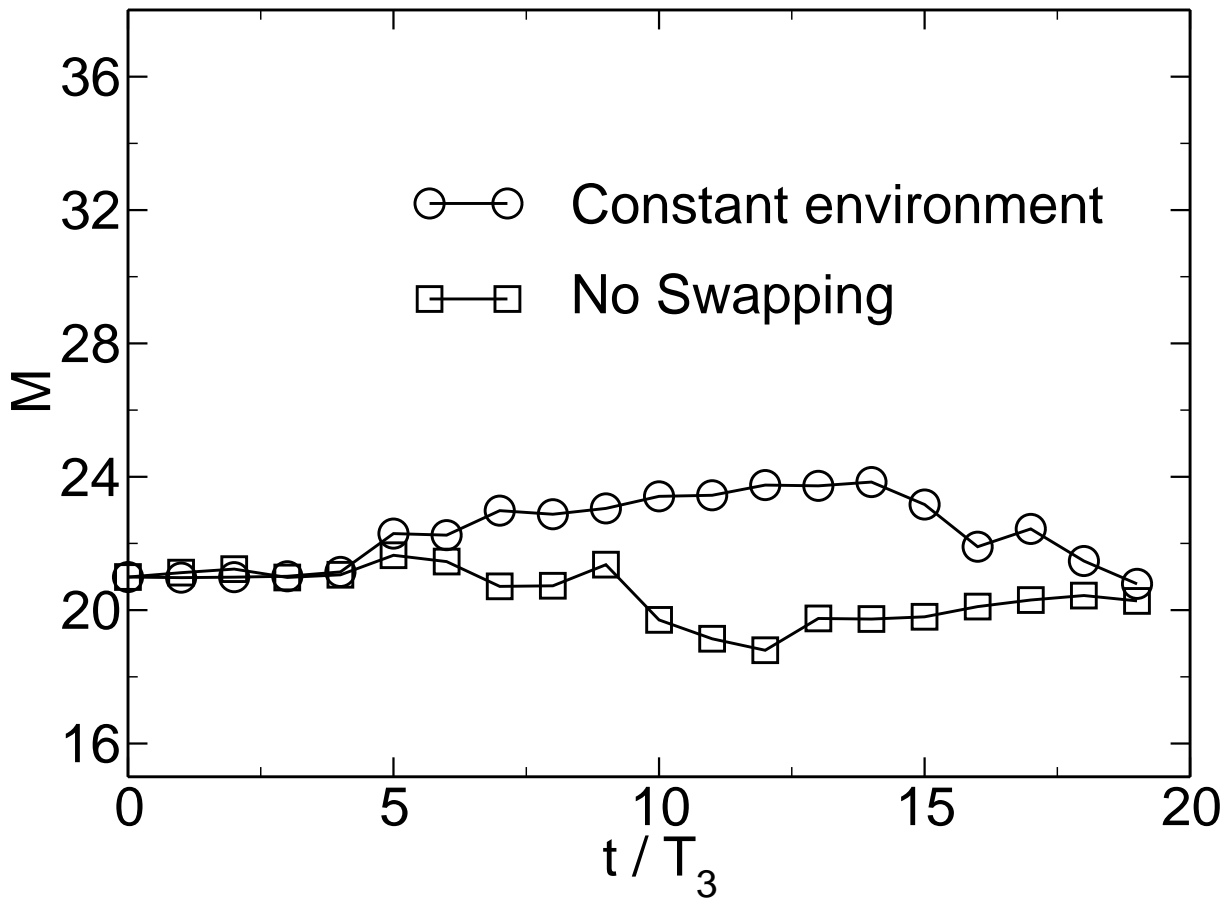}
\end{center}
\caption{Emergence of modularity in this model
requires both horizontal gene transfer and a changing environment.
Here $T_2=20$.
For the case of no horizontal gene transfer with a changing environment,
the severity of environmental change is $p=0.40$. 
For the case of no environmental change with horizontal gene
transfer, $p=0$, and the transfer is of fixed position $10 k + 1$ and fixed
length 10 and attempted every $T_2$ steps.
\label{fig:MandT-O-6}
}
\end{figure}

When the environment does not change, or if there
 is no horizontal gene transfer,
the modularity does not spontaneously emerge. As shown in 
Fig.\ \ref{fig:MandT-O-6},
the modularity remains constant at $M_0$
without environmental change or gene transfer.
\begin{figure}[t]
\begin{center}
\epsfig{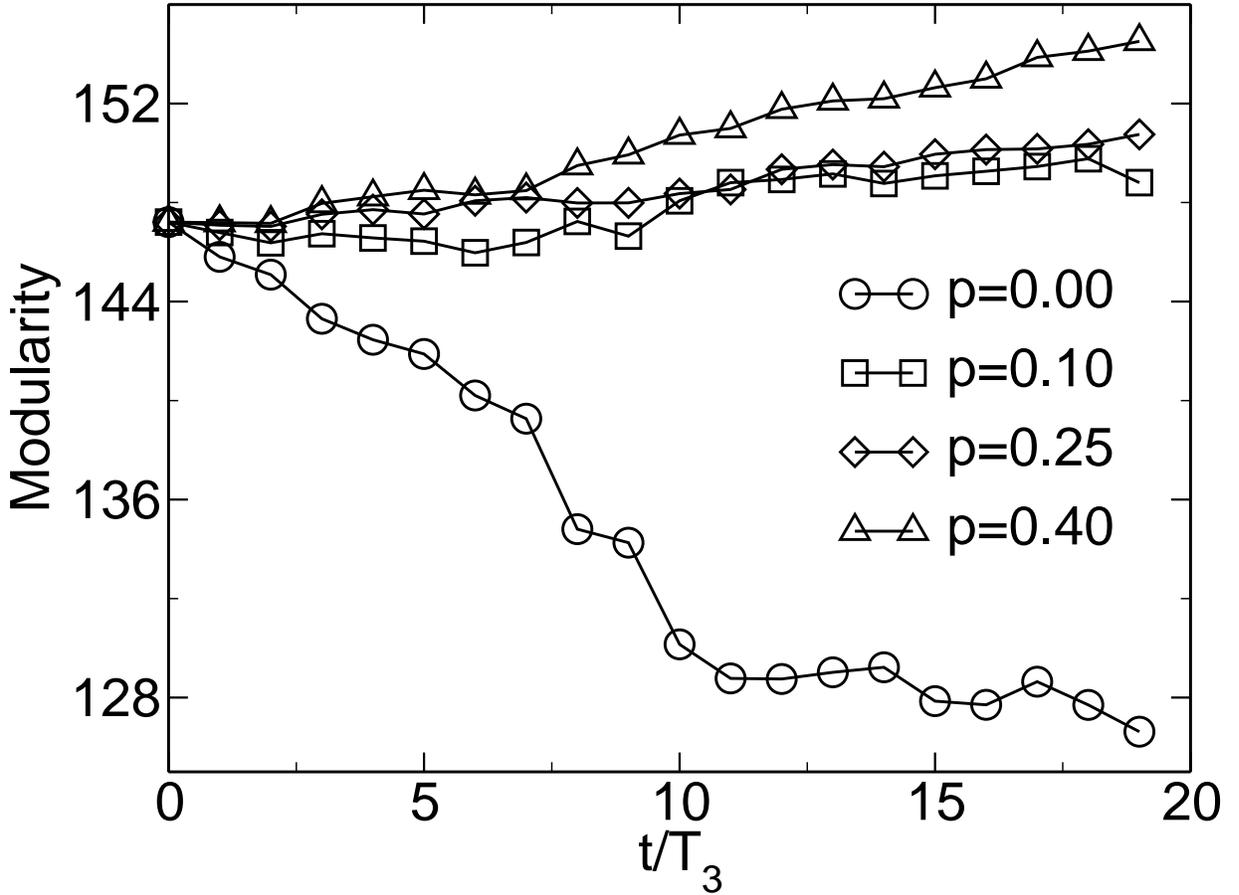}
\end{center}
\caption[]{The rate at which modularity grows, $dM/dt$, is positively
correlated with the magnitude of environment change, $p$. The frequency
of environment change is set at $1/T_2=1/40$.
\label{fig:MandT-A-0}
}
\end{figure}
\begin{figure}[t]
\begin{center}
\epsfig{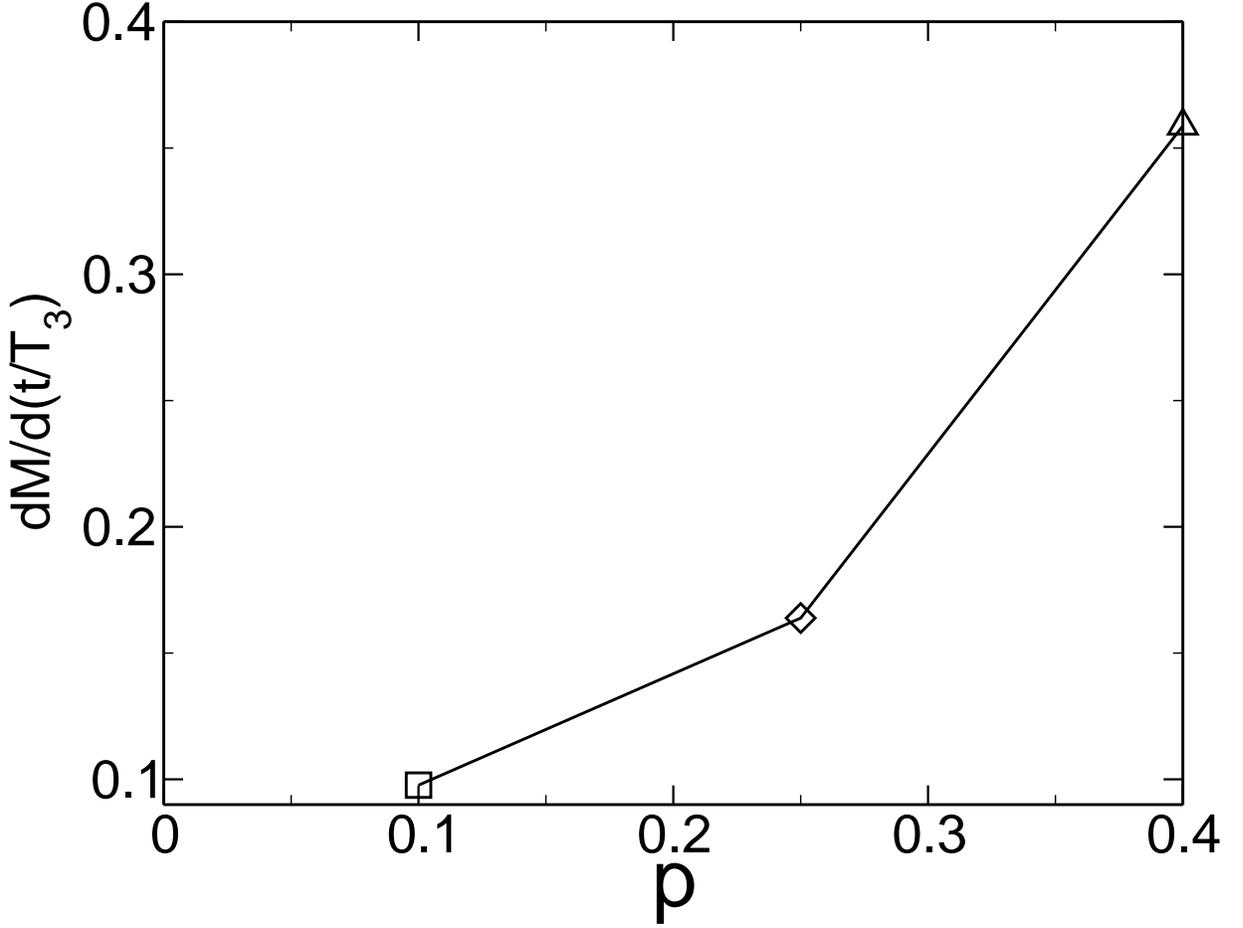}
\end{center}
\caption[]{
 The response function of the system, $dM/d (t/T_3)$, as a function of
the severity of environmental change for the data of
Fig.\ \ref{fig:MandT-A-0}.
\label{fig:MandT-A-1}
}
\end{figure}
The system adopts the broken-symmetry, modular state not because
the mutation and horizontal gene transfer moves favor modularity
a priori, but rather because these moves enable the system to respond
more effectively to a changing environment when the system is modular.
That is, evolvability is implicitly selected for in a changing environment, and
horizontal gene transfer enhances evolvability if the system is modular.
Thus, we expect modularity to be implicitly selected for in a changing
environment in the presence of horizontal gene transfer,
with the degree of modularity positively
correlated to the degree of environmental
change.
In Fig.~\ref{fig:MandT-A-0} we show the change of modularity with time
for different severities of environmental change, $p$.
For this figure, we choose the initial set of structures from
an ensemble with
$M=147$, rather than $M=M_0$, to show the change of
modularity more clearly.
For no environmental change, the modularity decreases
from this high level.  But for modest environmental change, the
modularity increases from the initial, high level.   
When the environment changes greatly,
the system must carry out more genetic change to survive, 
and it evolves a greater increase
of modularity. In Fig.~\ref{fig:MandT-A-1},
$dM/d(t/T_3)$ is the rate of the increase of modularity, and we show that the
rate of the
increase of modularity is larger for greater environmental change.

\begin{figure}[t]
\begin{center}
\epsfig{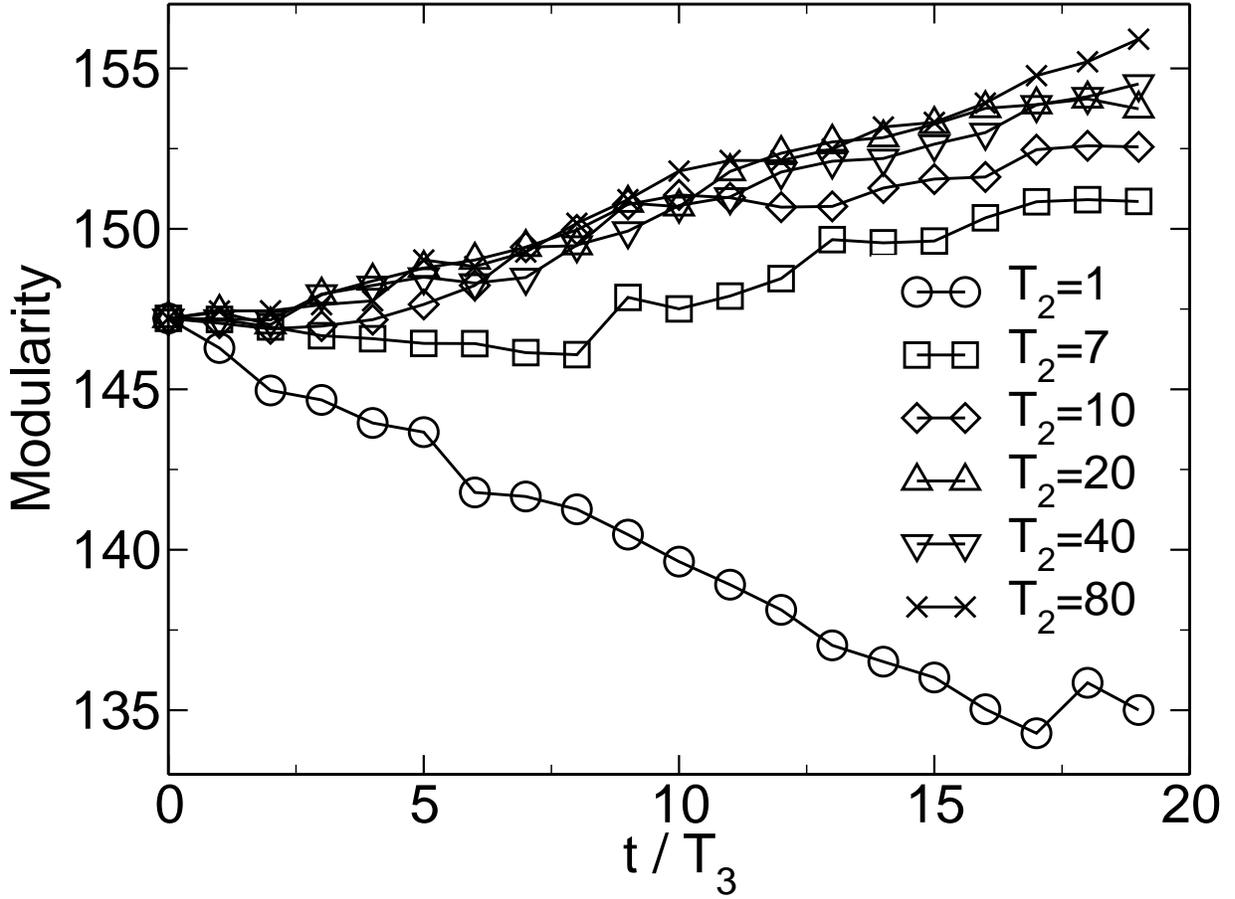}
\end{center}
\caption{Frequency of environmental change also affects the
time evolution of spontaneous modularity. 
Here $1/T_2$ is 
the frequency of environmental change, and
the severity of environment change is $p=0.40$.
}
\label{fig:MandT-F-0}
\end{figure}

\begin{figure}[t]
\begin{center}
\epsfig{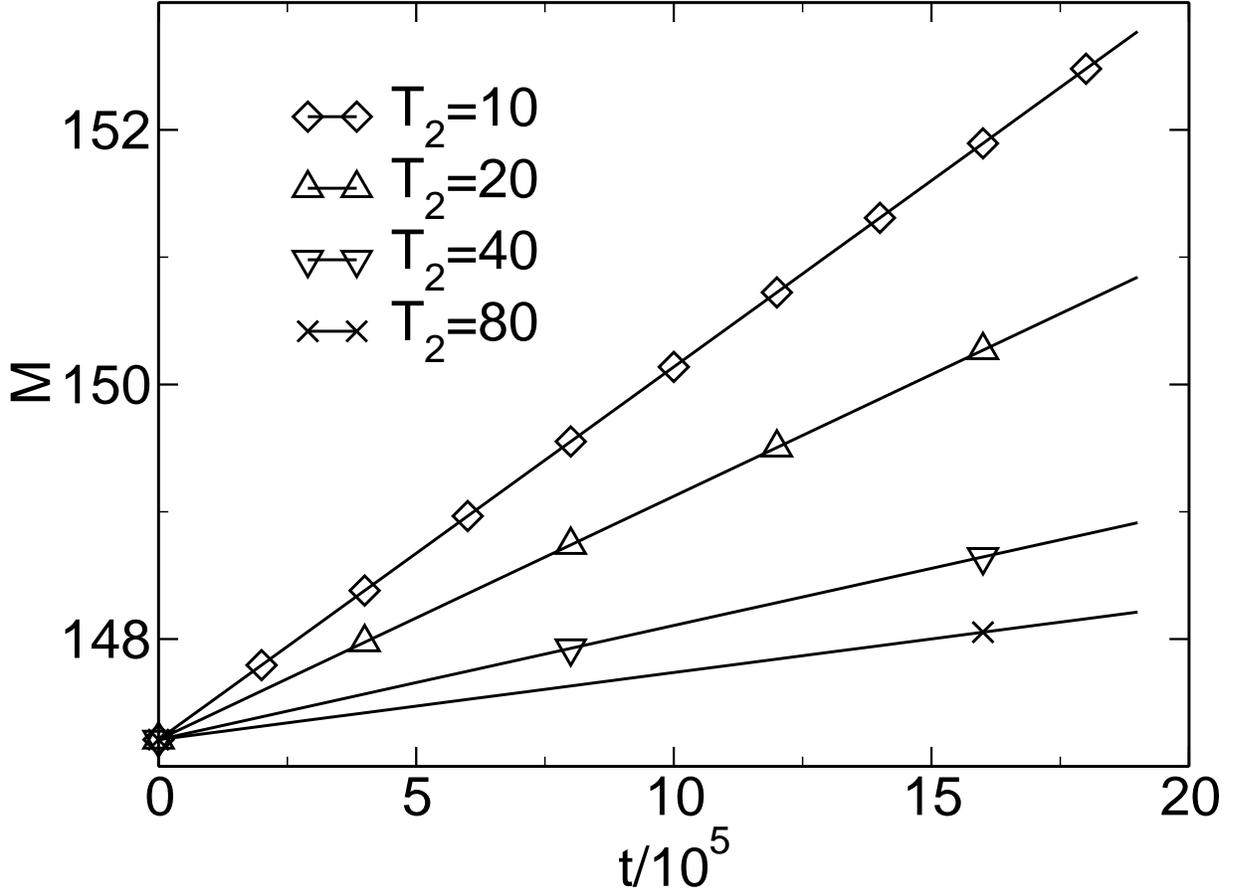}
\end{center}
\caption{Frequency of environmental change  affects the
time evolution of spontaneous modularity shown in real time, $t$.
Here $1/T_2$ is 
the frequency of environmental change, and
the severity of environment change is $p=0.40$.
\label{fig:MandT-F-1}
}
\end{figure}

\begin{figure}[t]
\begin{center}
\epsfig{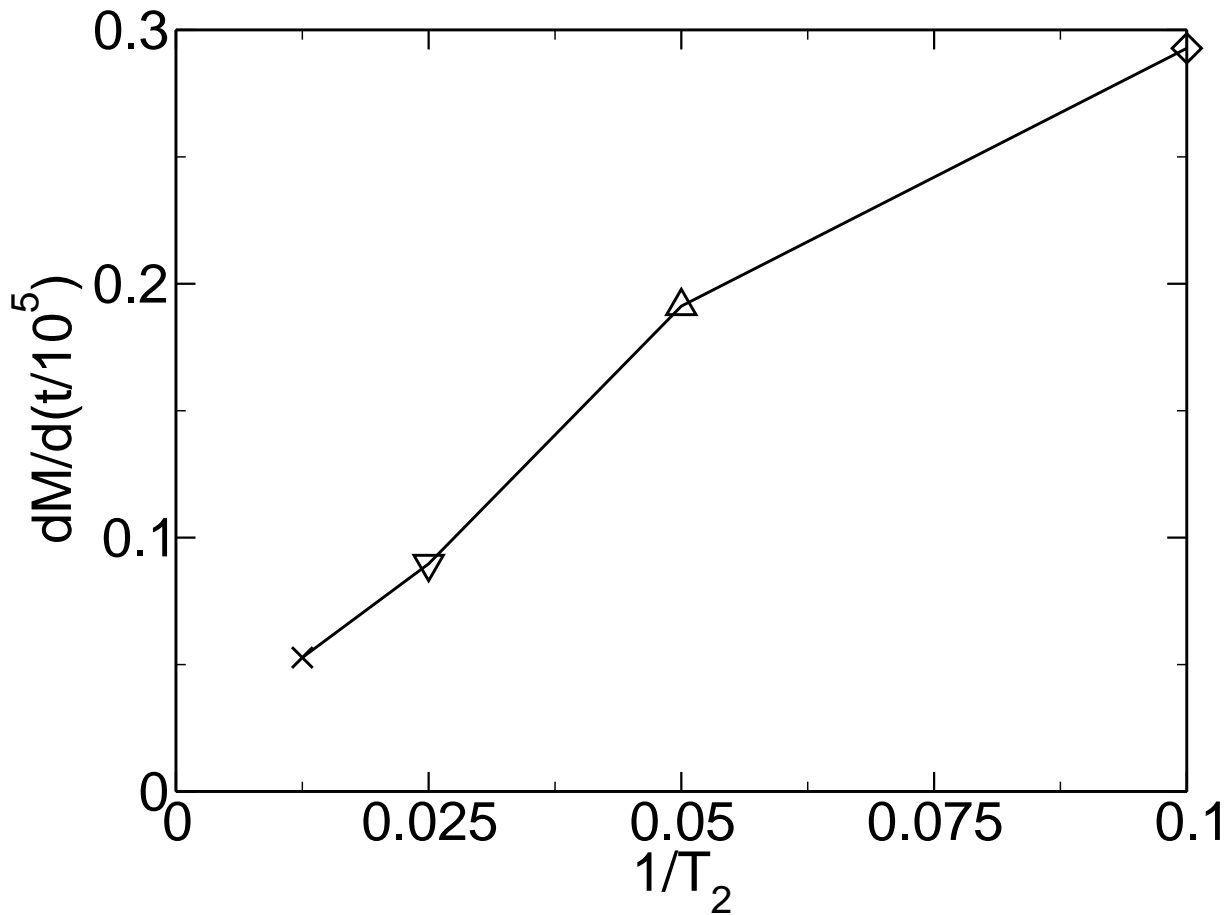}
\end{center}
\caption{The response function of the system $dM/d (t/10^5)$ as a function of
the frequency of environmental change ($1/T_2$) for the
data from Fig.~\ref{fig:MandT-F-0}.
\label{fig:MandT-F-2}
}
\end{figure}

Another way of characterizing the environmental change is by the
frequency of change, and the emergence of modularity depends on this
parameter as well.
In Fig.~\ref{fig:MandT-F-0} we show the growth of modularity with
time for different frequencies of environmental change.
For frequencies of environmental change that are not too large, the
modularity increases with frequency.  For very high frequencies,
$1/T_2 > 1/5$, the system is unable to track the changes in the environment,
and the modularity decays with frequency. Fig.~\ref{fig:MandT-F-1} is the same as Fig.~\ref{fig:MandT-F-0}
but with a real time as the $x$-axis. We can see that the increase of modularity is almost linear.
The rate of modularity increase in 
Fig.~\ref{fig:MandT-F-0} 
for $p=0.40$ and $T_2=20$ is less than that in
Fig.~\ref{fig:MandT-O-0} because in 
Fig.~\ref{fig:MandT-F-0}  the system is
closer to the steady-state, broken-symmetry value than it is in the
Fig.~\ref{fig:MandT-O-0}. In Fig.~\ref{fig:MandT-F-2}, we show that the 
rate of the increase of 
modularity is larger for higher frequencies of environmental change.

The spontaneous emergence of modularity is caused by the historical
variation in environments that the system has encountered.  By
a fluctuation-dissipation argument \cite{Hanngi1998,Kaneko,Earl},
 we might expect that
the degree of modularity should be proportional to the variance of
environments encountered.
In  Fig.\ \ref{fig:MandT-A-1} we show that the
rate of the increase in modularity is roughly proportional to the
severity of environmental change, $p$.
In Fig.\ \ref{fig:MandT-F-2} we show that the
rate of the increase in modularity is roughly proportional to the
frequency of environmental change, $1/T_2$.

\begin{figure}[t]
\begin{center}
\epsfig{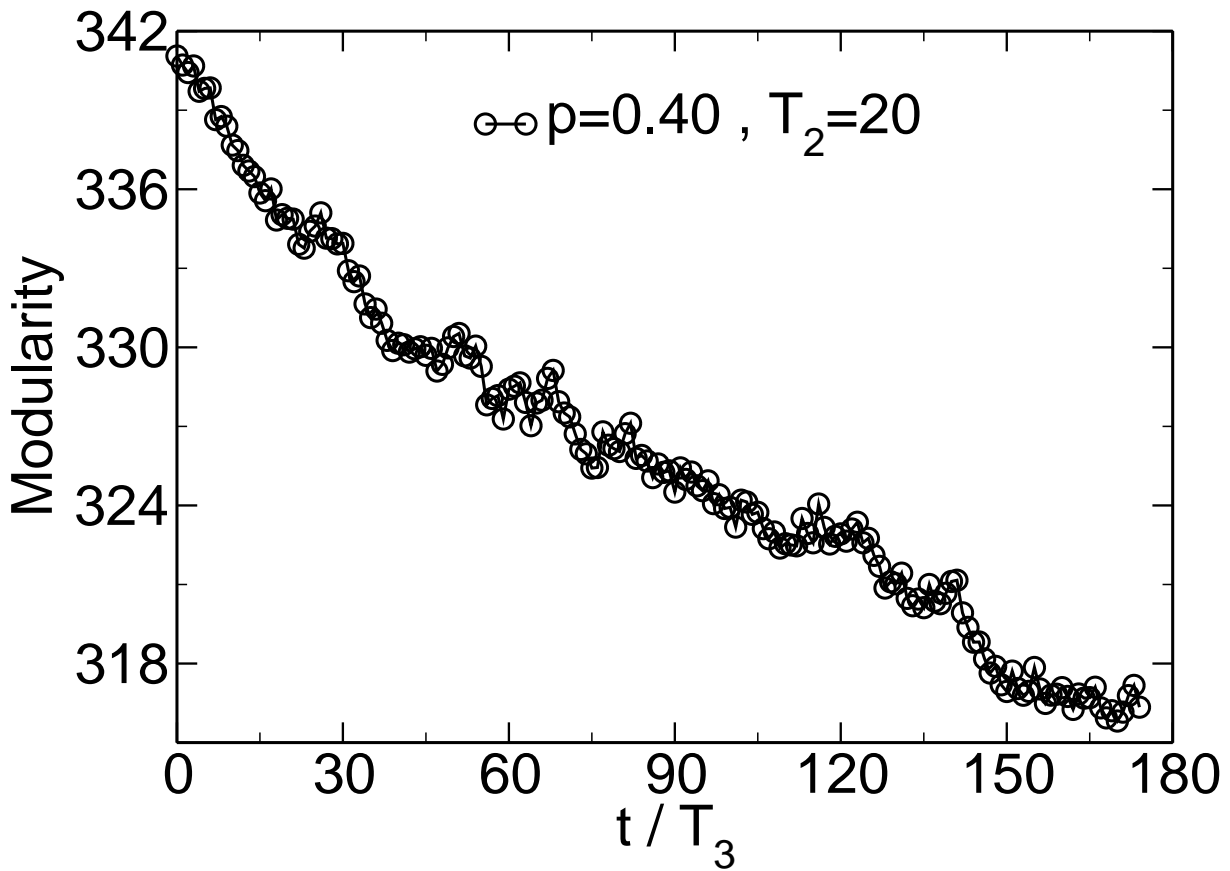}
\end{center}
\caption[]{
The spontaneous modularity saturates at a steady-state level.  
If the initial value of the modularity is greater than the
steady-state value, the modularity decays with time.
Here $T_2=20$, and the severity of environment change is 
$p=0.40$.  
}
 \label{fig:MandT-S}
\end{figure}
While the modularity grows with time in
Figs.\ \ref{fig:MandT-O-0},  \ref{fig:MandT-A-0}, and  \ref{fig:MandT-F-0}  for
$p>0$ and $T_2 >5$, at steady state the system will
be only partially modular, $M < N_D=346$, 
reflecting a balance between the selection
for modularity in a changing environment and 
the mutations driving the system toward the symmetric state
of no excess modularity.
To illustrate this point of a finite modularity in the steady state,
we show in Fig.~\ref{fig:MandT-S} how the modularity changes
from a starting point of nearly total modularity, $M \approx N_D$,
i.e.\ nearly all the connections in the diagonal blocks
and few in the off-diagonal blocks.
We observe that the modularity decays from the initial value.
See Fig.~\ref{fig:MandT-S}.
The excess modularity in the broken symmetry state is
positive because of
selection for modularity in fluctuating environments, and the
excess modularity is not the maximal possible value 
of $M=N_D=346$
because of the entropic effects of the mutations in the sequence space.
For the initial condition used in Fig.~\ref{fig:MandT-S},
nearly all the connections in the diagonal blocks
and few in the off-diagonal blocks,
modularity decays over time, showing the steady state value is
below $316$. The modularity will saturate at a value for which
the effects of selection pressure and 
mutation balance each other.

To summarize, we have observed the spontaneous
evolution of modularity in a population evolving in a changing
environment.  While we have described the model parameters in terms
of an evolving population of proteins, the model generically represents
evolution of individuals in a population with a non-trivial
microscopic fitness landscape.
Modularity arises spontaneously because evolvability is selected
for in a changing environment \cite{Earl}, and modularity allows
the horizontal gene transfer to rapidly evolve the system in a 
modified environment.  Thus, modularity is selected for in a changing
environment, when the system has access to horizontal gene transfer.
The rate at which modularity grows with time depends on the
amplitude and frequency of environment changes. More rapid
environmental change tends to promote the growth of modularity.
A constant environment promotes no emergence of modularity, as does
the limit of an extremely rapidly varying environments, because
the system sees only the average, constant environment.
The growth of modularity is also accelerated 
by more severe, larger-amplitude environmental changes.

\section{discussion}
In this section, we present some experimental evidence in
support of our simulation results.
The biological results  pertain to the specific instance
of our model as describing the formation of structure in
the protein-protein interaction network.
Parter \textit{et al.} \cite{parter2007}  found that 
bacteria with habitats in more variable environments have 
metabolic networks that are significantly more modular than do
bacteria with more constant habitats.
Kreimer \textit{et al.} \cite{kreimer2008} found
that bacteria inhabiting a greater number of niches have more
modular metabolic networks, and that horizontal gene transfer
contributed to modularity.
Singh \textit{et al.} \cite{singh2008}
found that stress response networks such as chemotaxis that
directly interact with the environment are more
modular than are 
stress response networks 
more insulated from the impact of environment,
such as competence for DNA uptake.
After reviewing these results,
we investigate the evolution of protein interaction network and 
protein domain interaction network in \textit{E.\ coli} and
\textit{S.\ cerevisiae}. We find that the modularity of 
both networks in both organisms appears to have
increased during evolution.

\subsection{Networks in variable environment are more modular}
In our simulation, we predicted that environmental change is a key factor of 
emergence of modularity \cite{jun}.
Networks in a severely changing environment are more modular.

Parter \textit{et al.} constructed the metabolic network of  117 bacterial
 species \cite{parter2007}. They normalized
the Newman modularity to allow comparison of the modularity of
networks with different size and degree \cite{parter2007}, and they calculated
the modularity of all the 117 bacteria species. They evaluated the variability
of environment by classifying the 117 bacterial species into 6 classes
according to the degree of variability of natural habitat. The 6 classes
in the order of increasing environmental change are: \textit{obligate
bacteria, specialized bacteria, aquatic bacteria, facultative bacteria, 
multiple bacteria, terrestrial bacteria}. They averaged the modularity of
bacterial species in each class, and found that networks in variable
environment are more modular than networks of species which evolved in
constant environment.

Kreimer \emph{et al.} investigated  metabolic networks
across the bacterial tree of life \cite{kreimer2008}.
They systematically calculated the Newman modularity for more than
300 bacterial species.  They found that bacteria occupying a limited
number of niches, such as endosymbionts and mammal-specific pathogens,
have metabolic networks that are less modular that are the metabolic
networks from species occupying a grater variety of niches.  In particular,
pathogens that alternate between hosts have more modular metabolic
networks than do single-host pathogens.  Finally, the degree of horizontal
gene transfer was positively correlated with the
modularity of metabolic networks.

Since the emergence of modularity is promoted by environmental
change, it is very likely that networks which directly
interact with environment are more modular than networks which are far
from the impact of environment. Singh \textit{et al.} \cite{singh2008}
reconstructed  three regulatory networks underlying stress response
(chemotaxis, competence for DNA uptake, and endospore formation) in
hundreds of bacterial and archaeal lineages. Chemotaxis is a canonical
signal transduction pathway which directly interacts with environment;
sporulation is closely tied to essential replication apparatus and is
strongly affected by the environment. Environmental change has great
selection pressure on these two networks. Conversely, competence for DNA uptake
has wide phyletic distribution and the impact of environment is limited.
Singh \textit{et al.} reported that chemotaxis networks display well modular
organization with  five coherent modules whose distribution among different
species shows great interdependence and rewiring. The sporulation network
is somewhat modularity, and the chemotaxis network is even more modular.
Conversely, competence for DNA 
uptake displays no modular structure. These results clearly support the
impact of environmental change on the emergence of modularity of stress
response networks.

\subsection{Modularity increases in protein networks and protein domain networks}

\subsubsection{A Definition of Compositional Age}
To study modularity in biology, we need both a
quantitative definition of modularity and a calibration of time of
divergence for the biological objects of interest. 
We here use the compositional age approach to quantify the divergence
time of a protein \cite{r7}.  In this method the order of
appearance of the amino acids over
time is identified, and an integer representing age of introduction
assigned to each amino acid.  The order is given \cite{r7} as
A/G=17, D/V=16, S=15, P=14, E/L=13, T=12, R=11, I=10, Q=9,
N=8, K=7, F=6, H=5, C=4, M=3, Y=2, and W=1.
The compositional age of a protein is the average of these values
over the sequence of the protein.  The compositional age of
a species is the average of the compositional age of all the
expressed proteins in that species.
Proteins that contain a greater fraction
of the oldest amino acids are then identified as arising earlier
than those proteins that contain a greater fraction of the newer amino acids.
By averaging the compositional age of each of the proteins in
a species, we determine the average time of divergence of that species.
In this paper we make this method quantitative, calibrating it upon
time points over the last 3.5 billion years.
This method does not require us to identify \emph{a priori} the
ancient species.

\begin{figure}
\begin{center}
\epsfig{file=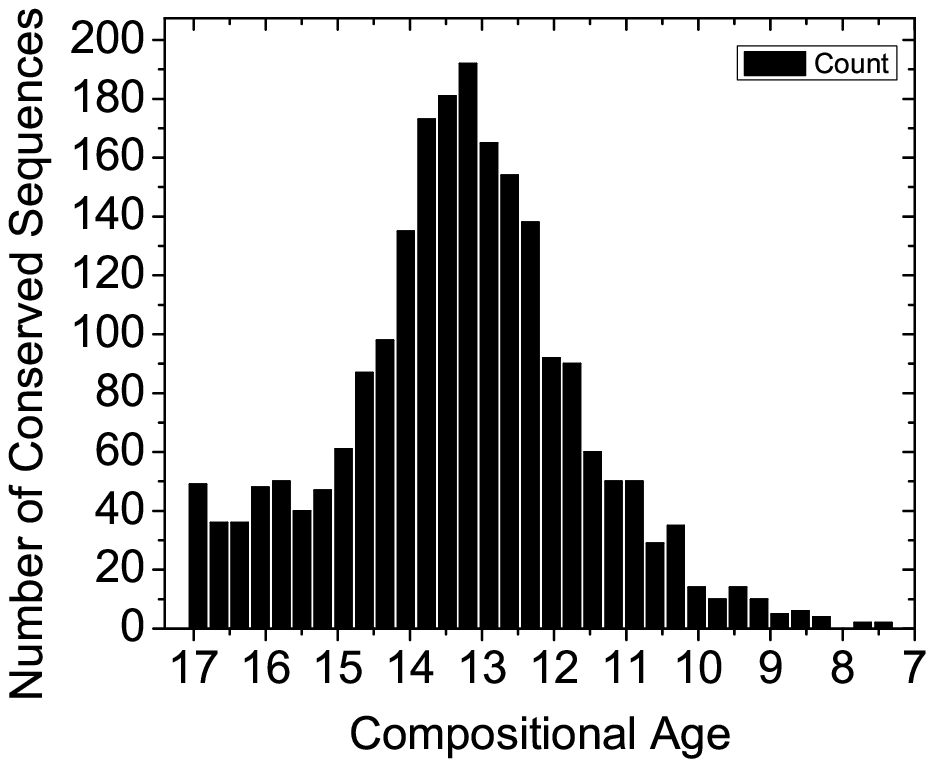,height=3.5in,clip=}\\
\hspace{0.9in} (a)\\
\epsfig{file=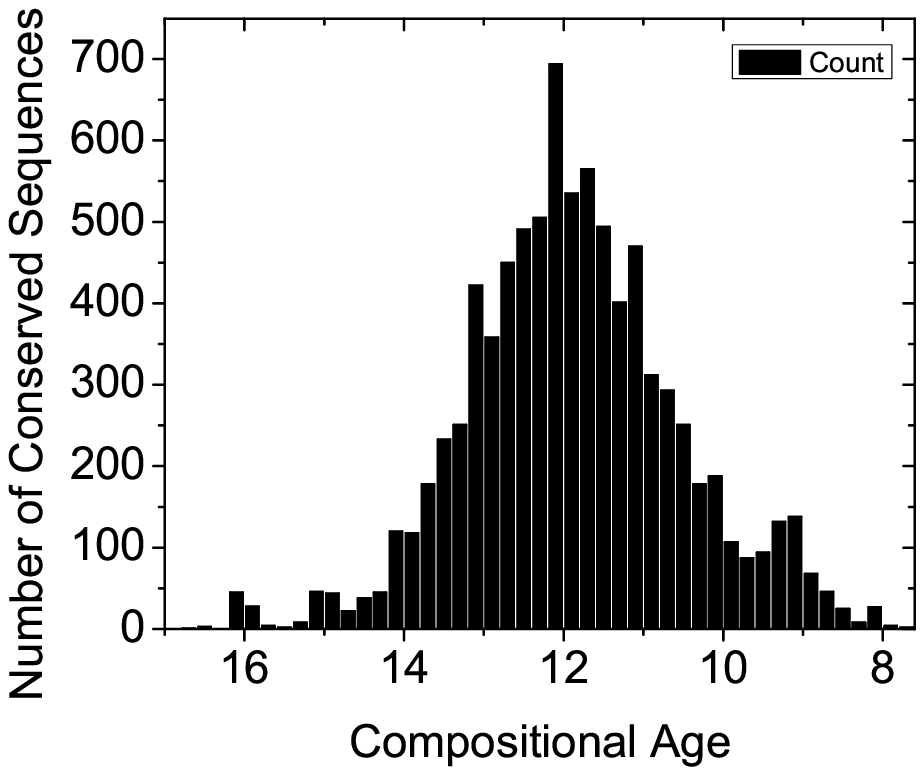,height=3.5in,clip=}\\
 \hspace{0.9in}(b) \\
\end{center}
\caption{ Distribution of conserved sequences with compositional age
to find (a) age of LUCA, and (b) divergence time of fungi.
 \label{age}
}
\end{figure}

To find the time of divergence of the earliest proteins,
we select 9 bacteria, 3 archaea, and 4 eukaryotic organisms to find the 
conserved sequences presumed to have arisen from
LUCA (Last 
Universal Common Ancestor). 
The bacterial species are 
\emph{ A.\ aeolicus, T.\ maritima, 
D.\ radiodurans, F.\ nucleatum, T.\ pallidum,
 C.\ glutamicum, C.\ acetobutylicum, S.\ aureus}, and \emph{ E.\ coli}.
The archaea species are
\emph{ A.\ fulgidus, S.\ solfataricus}, and \emph{ P.\ aerophilum}. 
The eukaryote species are \emph{ C.\ elegans, S.\ cerevisiae, S.\ pombe},
and \emph{ D.\ melanogaster}.  All the sequence data
come from EMBL-EBI. Using the software CONSERV
(http://www.gen-info.osaka-u.ac.jp/~ngoto/CONSERV/) 
we found 2163 conserved sequences with greater than 7 amino acids that appear 
in all the three kingdoms and in at least
8 proteins. We calculated the compositional age for these sequences.  A histogram
is shown in Fig.\ \ref{age}(a).
The distribution of compositional age peaks at 13.32. 
There is some debate about the age of LUCA, 
with estimates ranging from 3.5 to 4.0
 billion years ago \cite{r1}. In our work, we set LUCA
at the average of
3.8 billion years ago. Thus, we assign a compositional age of 13.32 
to a real age of 3.8 billion years ago.

To find the divergence times of fungal proteins,
we investigate 10 species of fungi.  
In the group Dikarya/Ascomycota/Saccharomycotina
we choose \emph{ S.\ cerevisiae,
C.\ glabrata, K.\ lactis, Y.\ lipolytica}, and \emph{P.\ stipitis}. 
In the group Dikarya/Ascomycota/Pezizomycotina we choose
 \emph{N.\ crassa,  M.\ grisea}, and \emph{ A.\ fumigatus}. 
We find 8535 sequences 
with greater than 15 amino acids that appear in both branches 
and in at least 4 proteins.  
The histogram of compositional age of these sequences
is shown in Fig.\ \ref{age}(b).
The compositional age peaks at 12.1. 
We choose $1.1$ billion 
years ago as the real age of divergence
time of these two branches of fungi \cite{r1}. So, the
compositional age of 12.1 is corresponds to an age of 1.1 billion years ago.

To find the compositional age of recent proteins,
we search for the youngest proteins in \emph{E.\ coli}. 
We consider only proteins in the COG (Clusters of Orthologous Groups
of proteins) database,
to exclude those protein fragment without 
function in the FASTA file.
 We compare the proteins in two strains of \emph{E.\ coli}:
K12 and o157 :H7 EDL 933. The 0157 strain of \emph{E.\ coli} diverged from
K12 strain about 4 million years ago \cite{Reid}.
We take the strains of \emph{E.\ coli} from the COG database that
exclude
the orthologous proteins that are shared by K12 and O157, which
should be quite young, probably less than $4$
million years. The youngest new protein of O157  has
compositional age of $9.607$. The youngest new protein of K12  
has compositional age of $9.652$.  We, therefore,
set the compositional age of present day as $9.6$.

\subsubsection{ Compositional Age and Evolutionary Rate}
\begin{figure}
\begin{center}
\epsfig{file=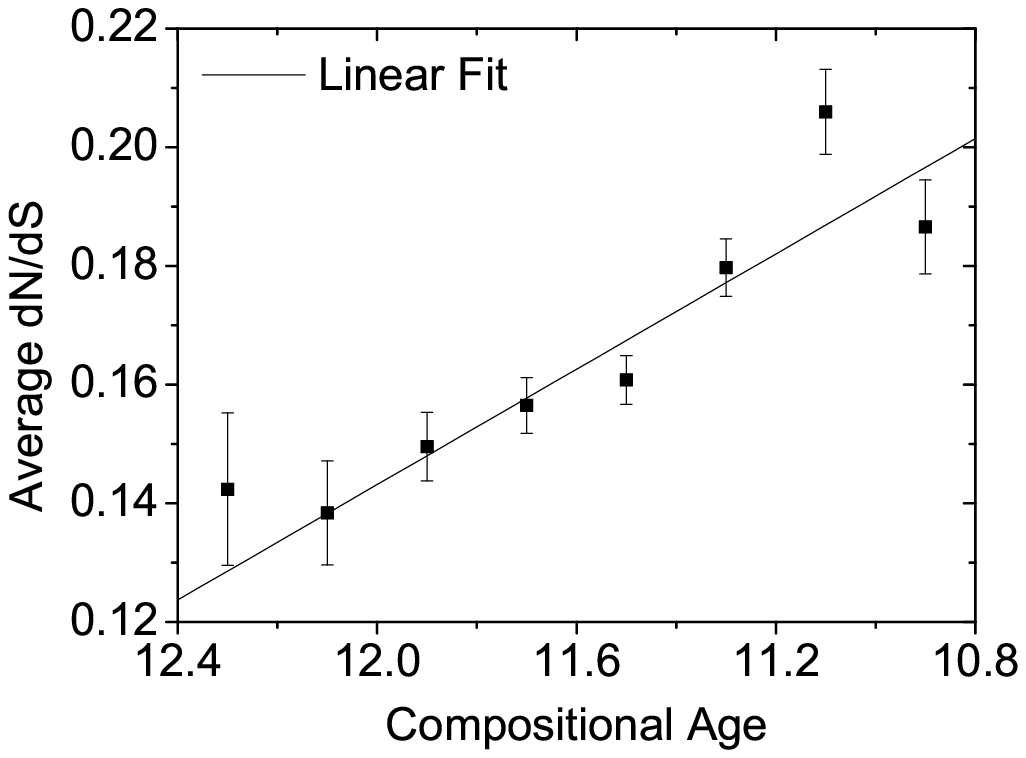,height=3.5in,clip=}\\
\end{center}
\caption{\emph{S.\ cerevisiae.} The average $dN/dS$ is negatively linearly
related to the compositional age.
\label{evol_age}
}
\end{figure}

We want to find the relationship between the
compositional age of proteins and the
evolutionary rate of corresponding genes.
The ratio of nonsynonymous substitution per site to synonymous
substitution per site ($dN/dS$)
is often assumed to be
a good measure of evolutionary rate. 
Hirsh \emph{et al.}\ compared the orthologous open reading frames in four
yeast spices and provided $dN/dS$ data for 3392 genes \cite{hirsh}. Here we
average the $dN/dS$ of proteins in every compositional age interval 0.2. 
For example,  there are 326 proteins with compositional age between 10.9 and 
11.1, we calculate the average $dN/dS$ 0.21 for those proteins and plot it in 
Fig.\ \ref{evol_age}.  The compositional age 
is negatively and nearly linearly related to the evolutionary rate 
(correlation coefficient $R^2 =0.83$).
 
\subsubsection{ Growth of Modularity in the Protein-Protein Interaction Network}
\begin{figure}
\begin{center}
\epsfig{file=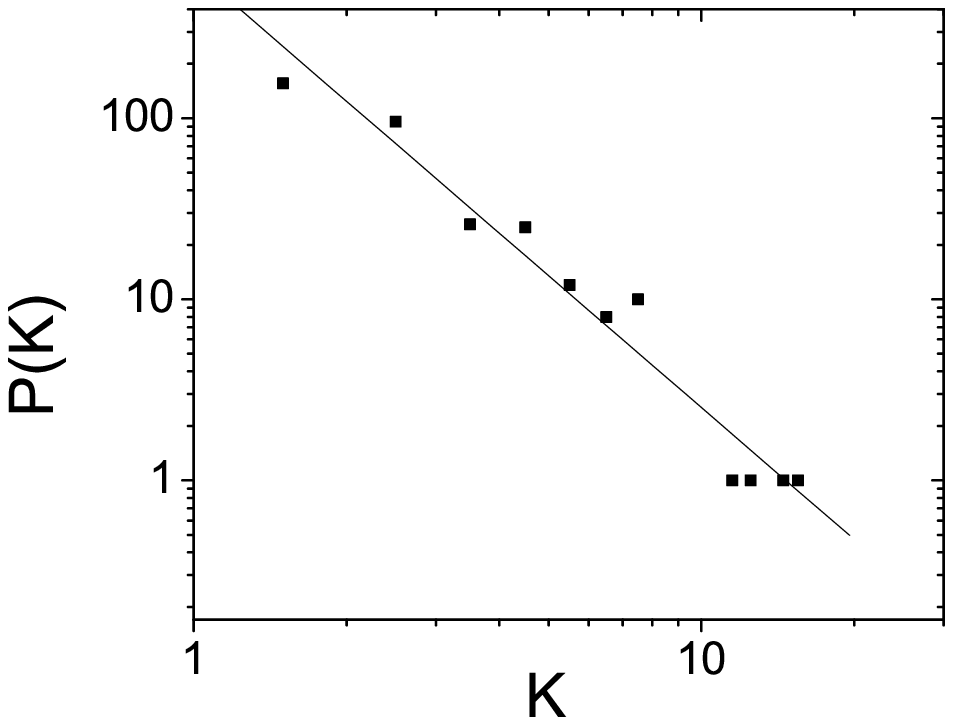,height=3.5in,clip=}\\
\end{center}
\caption{ The degree distribution of the \emph{S.\ cerevisiae} domain-domain
interaction network.
  \label{scale}
}
\end{figure}

We quantify modularity of both protein domain structure and of the protein-protein
interaction network \cite{r16,r23,r24}.  
The protein-protein interaction network data come from DIP.
We obtain 1846 proteins with 6971 interaction
edges in \emph{E.\ coli} and 3211 proteins with 17535 interaction edges in
\emph{S.\ cerevisiae}.
The domain-domain interaction data come from InterDom.
We consider only domain interactions
based on the DIP database and take only these domain interactions with a
score in the top 75\%, to eliminate the noisy data.
We obtain 276 proteins in \emph{E.\ coli} and 427
proteins in \emph{S.\ cerevisiae}, from which we
extract the protein domains for study.
Interestingly, the domain-domain interaction
network is scale free with $\gamma=2.4$, see Fig.\ \ref{scale}.

To quantify modularity in the
interaction networks, we construct the 
topological overlap matrix \cite{r26} from the interaction network,
reorder it with the
average linkage clustering method \cite{Botstein}, and normalize the
number of interactions within modules according
to network size.
The topological overlap matrix element, $T_{ij}$,
is the ratio of common nearest neighbors of the interacting
proteins $i$ and $j$ to their respective degrees.
The topological overlap matrix reflects the topological overlap
of the nearest neighbors of two nodes. For any two nodes $i$ and $j$,
the topological overlap is defined as \cite{r26}:
\begin{equation}
T_{ij}=\frac{\sum_{u}
a_{iu}a_{uj}+a_{ij}}{\min(k_i,k_j)+1-a_{ij}}
\end{equation}
Here $a_{ij}$ is the
elements of the interaction network matrix with value 0
(not interacting) or 1 (interacting).
We use the average-linkage
hierarchical clustering algorithm \cite{r26} to reorder the
topological overlap matrix so that the more tightly linked and
clustered nodes are moved close to each other. 
In this way,
we identify the modules and hierarchical structure of the network. 

\begin{figure}
\begin{center}
\includegraphics[width=.4\textwidth]{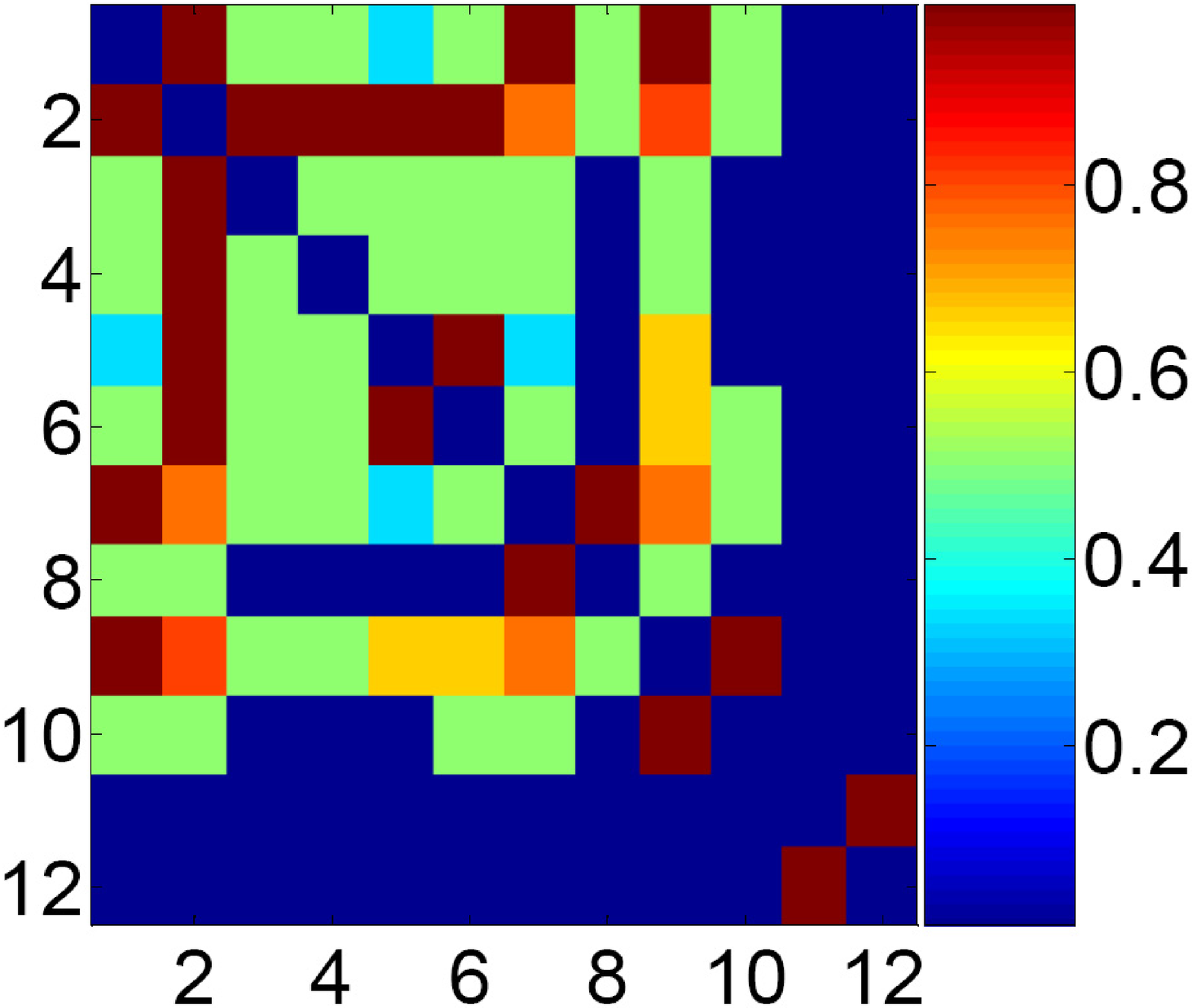}
\includegraphics[width=.4\textwidth]{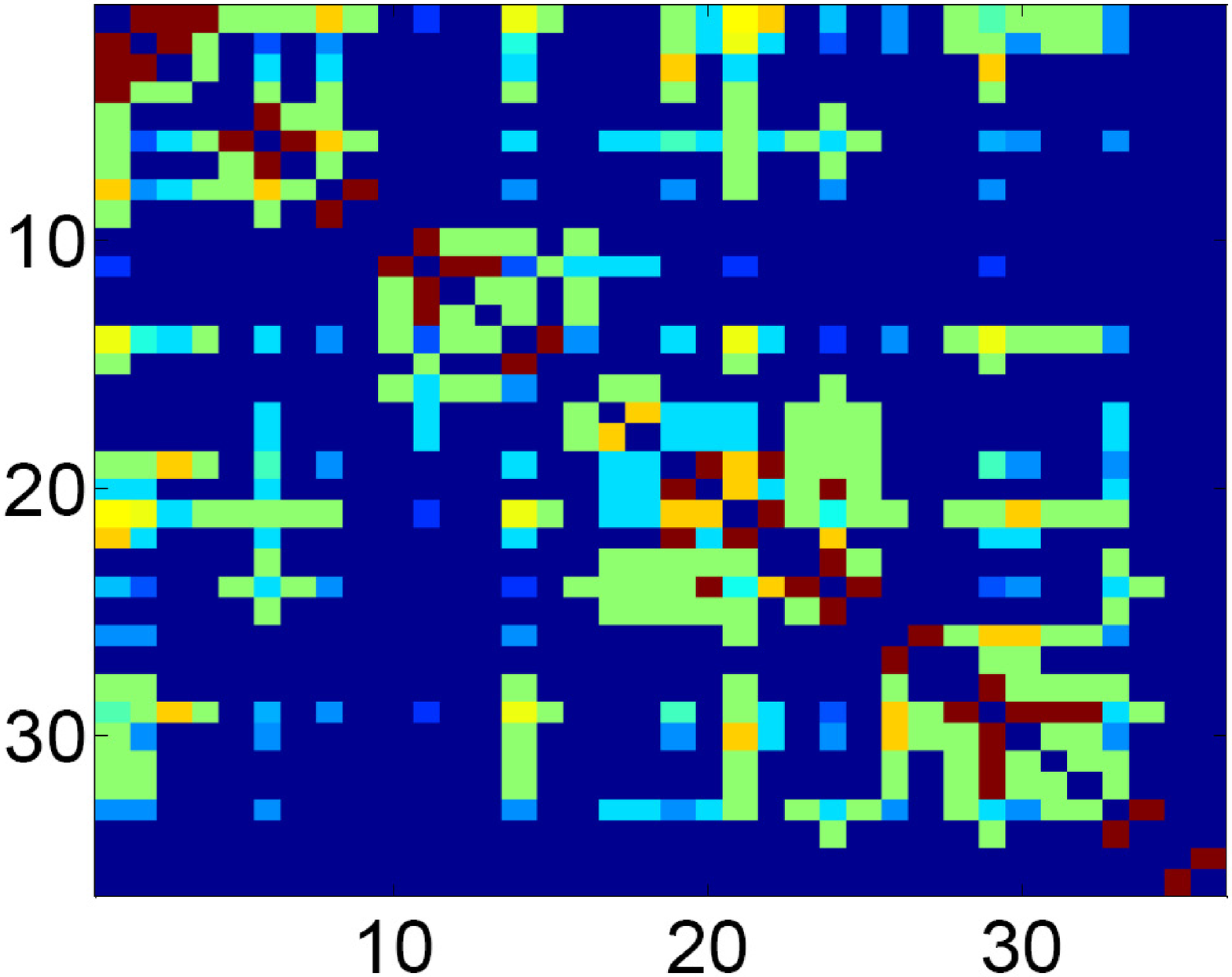}\\
\hspace{0.3in} (a) \hspace{2.5in} (b) \\
\includegraphics[width=.4\textwidth]{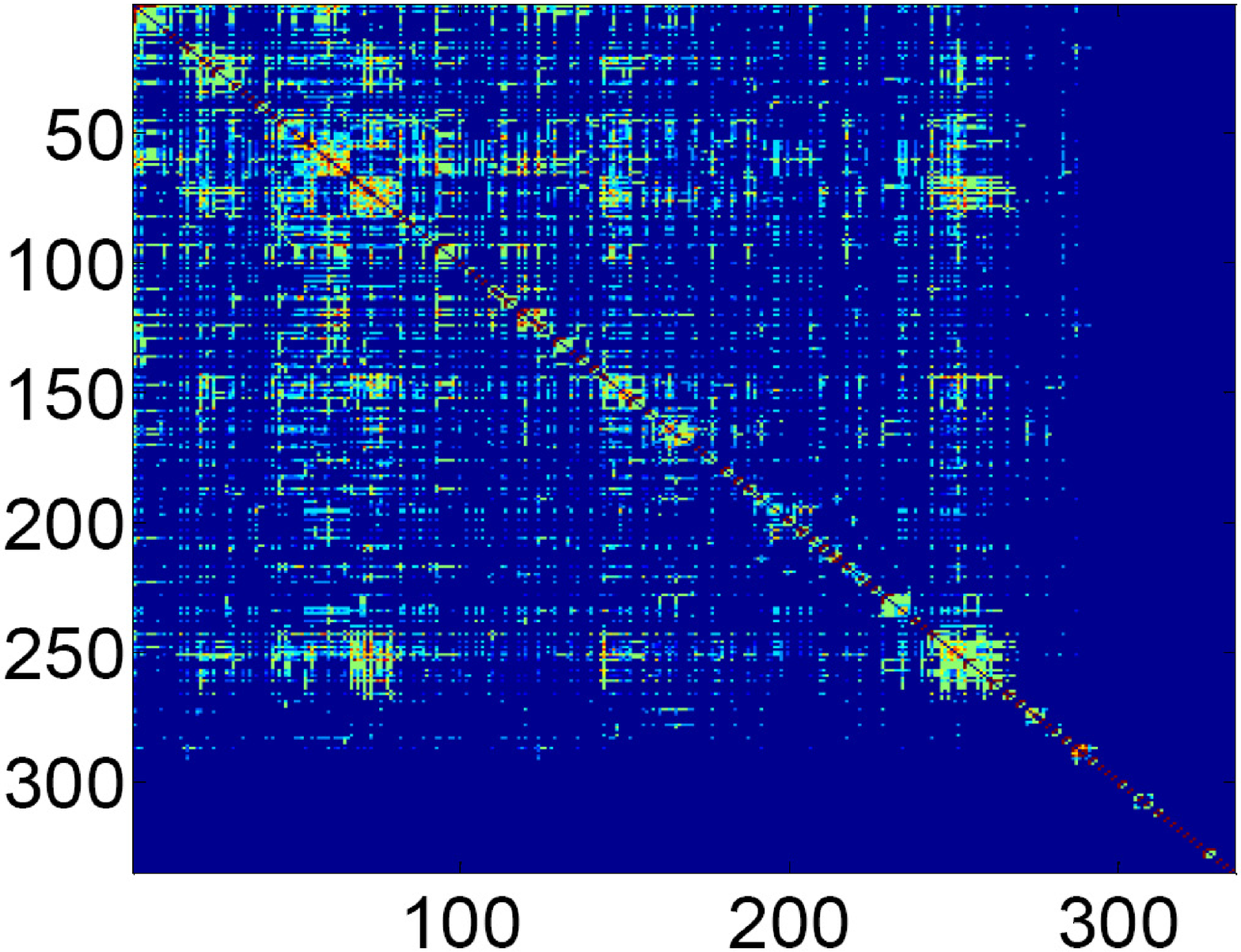}
\includegraphics[width=.4\textwidth]{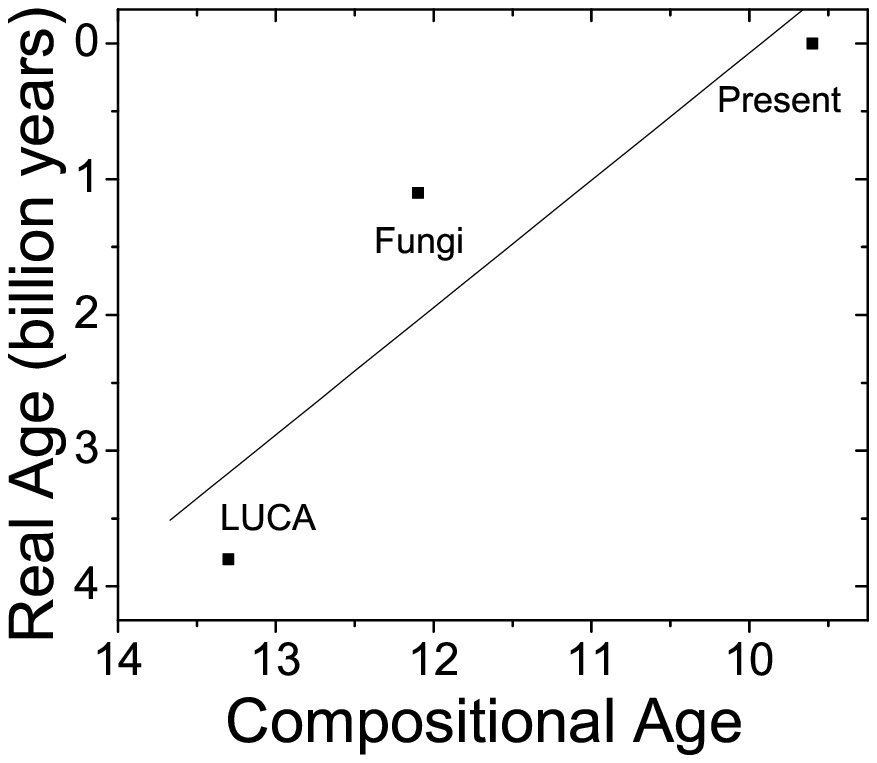}\\
 \hspace{0.3in} (c) \hspace{2.5in} (d) \\
\includegraphics[width=.4\textwidth]{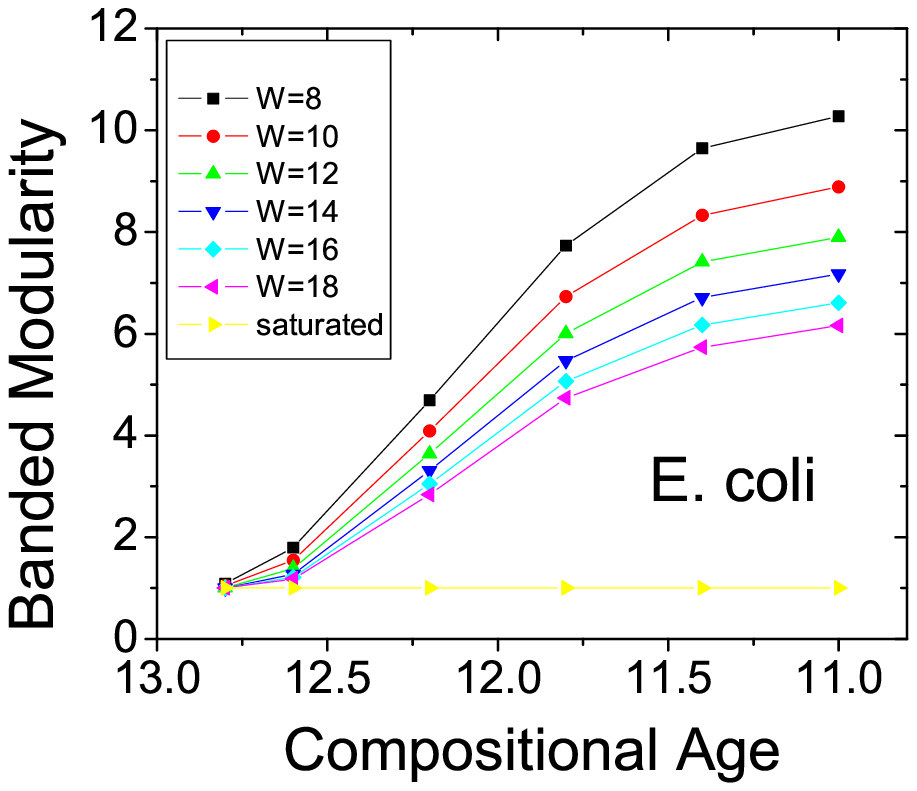}
\includegraphics[width=.4\textwidth]{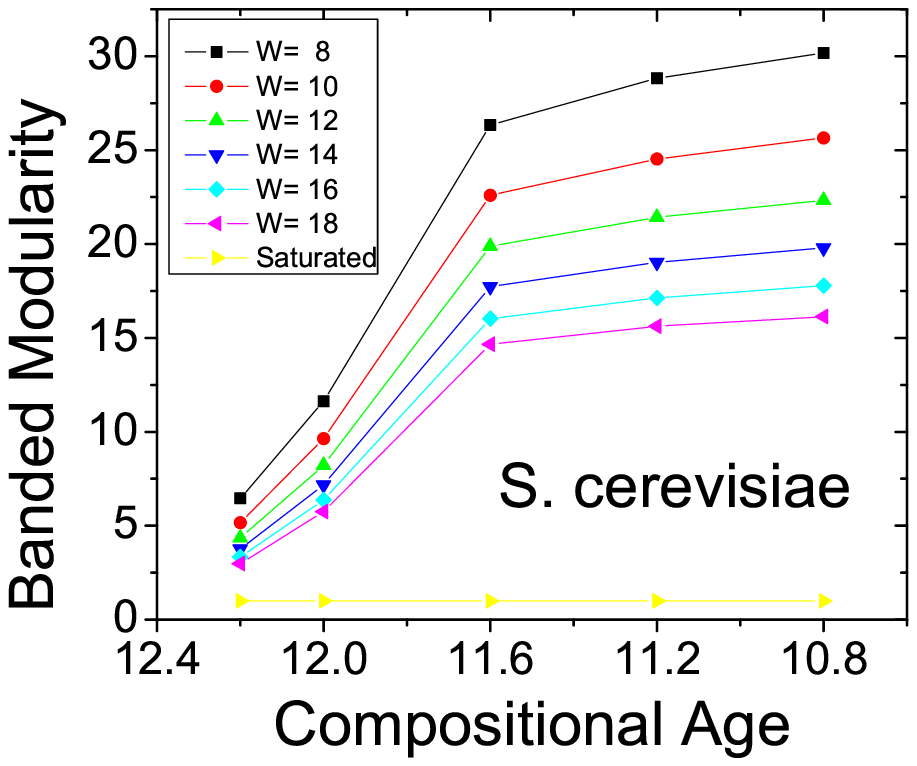}\\
 \hspace{0.3in} (e) \hspace{2.5in} (f) \\
\end{center}
\caption{(color online) 
The reordered topological overlap matrix of the \emph{E.\ coli} protein
interaction network constructed from proteins whose compositional age are
larger than 12.8 (a), 12.6 (b), and 12.2 (c). The color reflects the
strength of the topological overlap of two nodes (from 0.0 to 1.0),
as shown in the color bar in (a). (d) The linear relationship
between compositional age and real age.
(e) and (f), The banded modularity evolution of \emph{E.\ coli} and \emph{S.\ cerevisiae}, respectively. 
The lines of different color in (e)
and (f) correspond to different band sizes ($W$). 
Modularity grows with time.
Banded modularity of
a saturated matrix, \emph{i.e.}, a matrix with all elements being
$1$ except the diagonal ones being $0$, is shown in (e) and (f) for
comparison. The banded modularity of a saturated network is at its minimum
value of $1$. 
\label{ecoli_m}}
\end{figure}

The reordered topological overlap matrix of \emph{E.\ coli} at
different times is shown in Fig.\ \ref{ecoli_m}.
The protein-protein interaction network evolves from 
an almost saturated, unstructured network in
Fig.\ \ref{ecoli_m}(a) to a mildly modular network with four modules in
Fig.\ \ref{ecoli_m}(b) and then to a highly modular network in
Fig.\ \ref{ecoli_m}(c). To compare the modularity quantitatively, we
define banded modularity as the ratio of interaction within a diagonal
band to the total interactions, normalized by the ratio of the area of
the band to the area of the matrix: 
\begin{equation}
M_{\rm banded} = \frac{\sum_{0<\vert i-j \vert <W}^D
T_{ij}}{\sum_{i\neq j}^D
 T_{ij}}\times(\frac{\sum_{0<\vert i-j \vert <W}^D 1}{\sum_{i\neq j}^D 1})^{-1}
\end{equation}
Here, $W$ is the width of the band, $D$ is the dimension of matrix and
$T_{ij}$ is the element of reordered topological overlap matrix.
Since the network size grows in time, we compare modularity
of network of different sizes. The factor
$1/({\sum_{0<|i-j|<W}^D 1}/{\sum_{i\neq j}^D 1})$ normalizes for
the network size. 
In Fig.\ \ref{ecoli_m}(e), we show the banded modularity grows with compositional age in
\emph{E.\ coli}. A similar result is observed for \emph{S.\ cerevisiae} in
 Fig.\ \ref{ecoli_m}(f). 
This result holds true for different band widths
and different organisms; this phenomenon is robustly observed.
In a modular structure, there are more interactions within a module
than between modules. Banded modularity is a concise definition of modularity,
but may also be interpreted as simply locality, in which true modules
may not be identifiable. 

To measure modularity in a more detailed way, we search
along the diagonal of the reordered topological overlap matrix to
find the explicit modules, and we calculate the ratio of interactions
in the modules to the total interactions, normalized by the
ratio of the area of
modules to the area of the whole matrix.  
 We define these modules quantitatively.
First, we suppose the protein $i$ and $i+1$ form a module, and we
ask whether another another protein $i+2$,
should be added to the module.  We add the protein
if the average interaction between $i+2$ and
the existing module is larger than a
cutoff, which we set it 0.2 in our study.
We continue this procedure.
When we come to a protein with average interaction less than
the cutoff, this protein forms the first member of a new
module, and we begin the search to add further proteins to this new module.
The modules so identified depend on the cutoff.
In our study, the \emph{E.\ coli} and \emph{S.\ cerevisiae} networks
are highly modular. We
tried several cutoff and found the results are quite stable, with
results in accord with our visual observation of the clustered matrix.
We define the result as block
modularity: 
\begin{equation}M_{\rm module} = \frac{ \sum_{j,k\neq j=1}^{'D} T_{jk}
}{\sum_{j,k\neq j=1}^D T_{jk} } \times (\frac{
\sum_{j,k\neq j=1}^{'D} 1
}
{D(D-1)})^{-1}
\end{equation}
 where in the upper sum with the prime, $k$ is over
those proteins in the same module as $j$,
$D$ is the dimension of the matrix.

\begin{figure}
\begin{center}
\includegraphics[width=.4\textwidth]{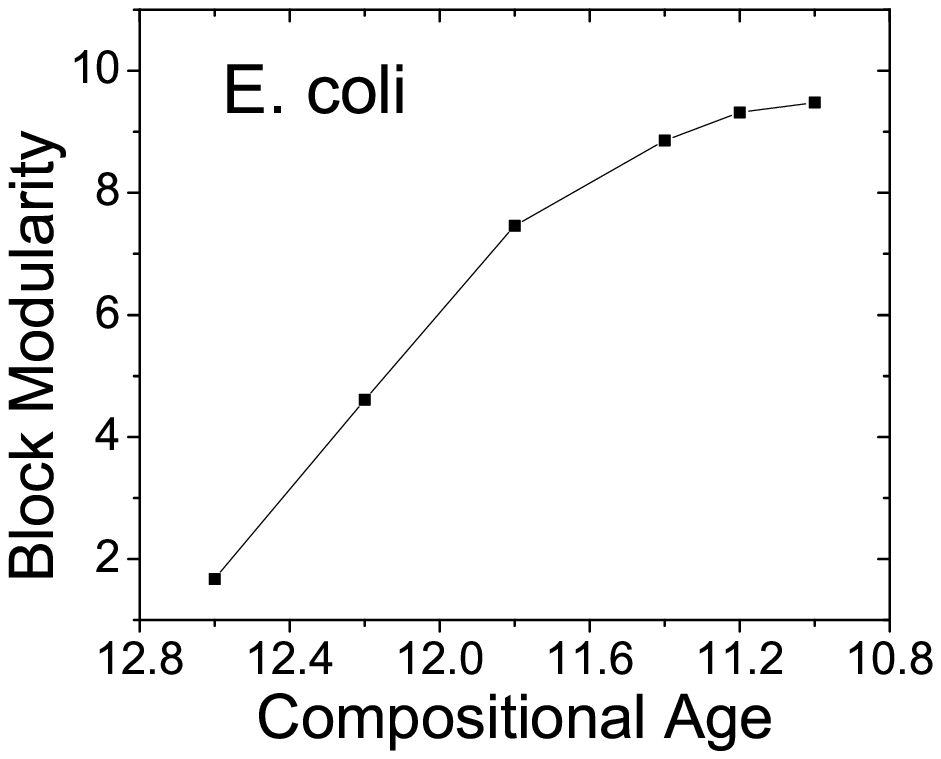}
\includegraphics[width=.4\textwidth]{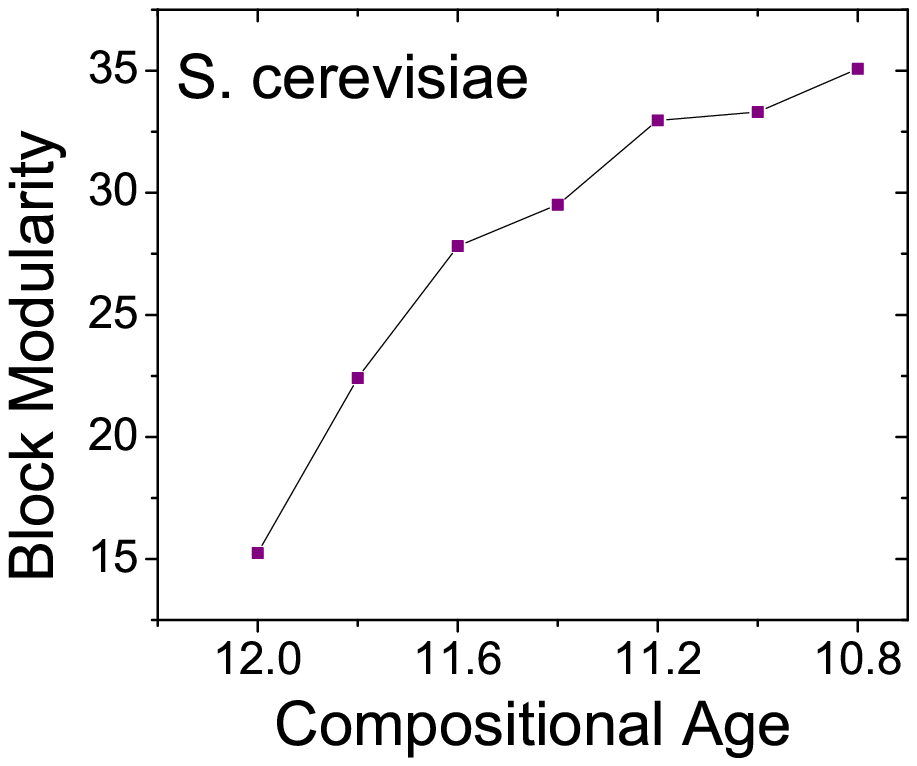}\\
\hspace{0.5in} (a) \hspace{2.2in} (b) \\
\end{center}
\caption[]{Evolution of block modularity of protein interaction
network in \emph{E.\ coli} (a) and \emph{S.\ cerevisiae} (b).
 \label{module}
}
\end{figure}

\begin{figure}
\begin{center}
\epsfig{file=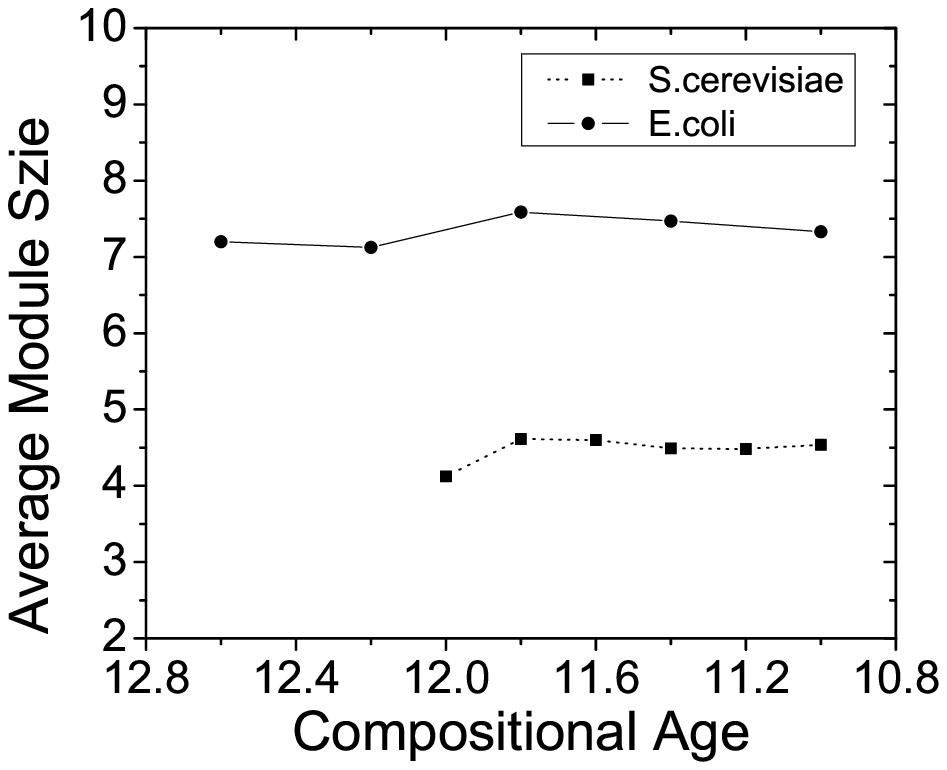,height=2.5in,clip=}
\epsfig{file=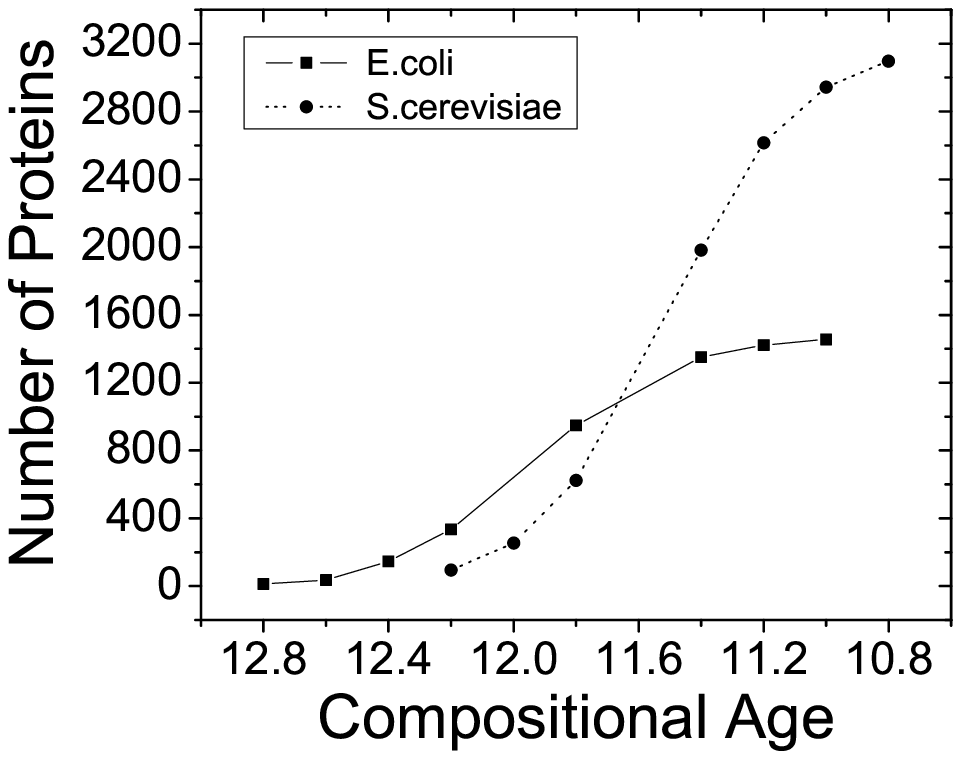,height=2.5in,clip=}\\
\hspace{0.3in} (a) \hspace{2.3in}(b) \\
\end{center}
\caption{ (a) Average number of proteins in a module
at different compositional ages,
 (b) Size of network in different compositional age network.
\label{module_size}
}
\end{figure}

We apply this definition to the reordered topological
overlap matrix to obtain the result for \emph{E.\ coli} and \emph{S.\ cerevisiae} in
Fig.\ \ref{module}. We see the growth of block modularity in both
organisms.  There is a positive correlation between banded and
block modularity. The growth of modularity is robust to the precise
definition of modularity.
The average size of module
at different compositional age network is stable, see
Fig.\ \ref{module_size}(a). The relationship between the size of the
network and compositional age is show in Fig.\ \ref{module_size}(b).
The average module size does not change much in evolution,
and the number of proteins in each module in
of \emph{S.\ cerevisiae} is fewer than that in \emph{E.\ coli}, perhaps
reflecting that \emph{S.\ cerevisiae} is more modular.

%
%
%

\subsubsection{ Growth of Modularity in the Domain-Domain Interaction Network}
\begin{figure}
\begin{center}
\includegraphics[width=.4\textwidth]{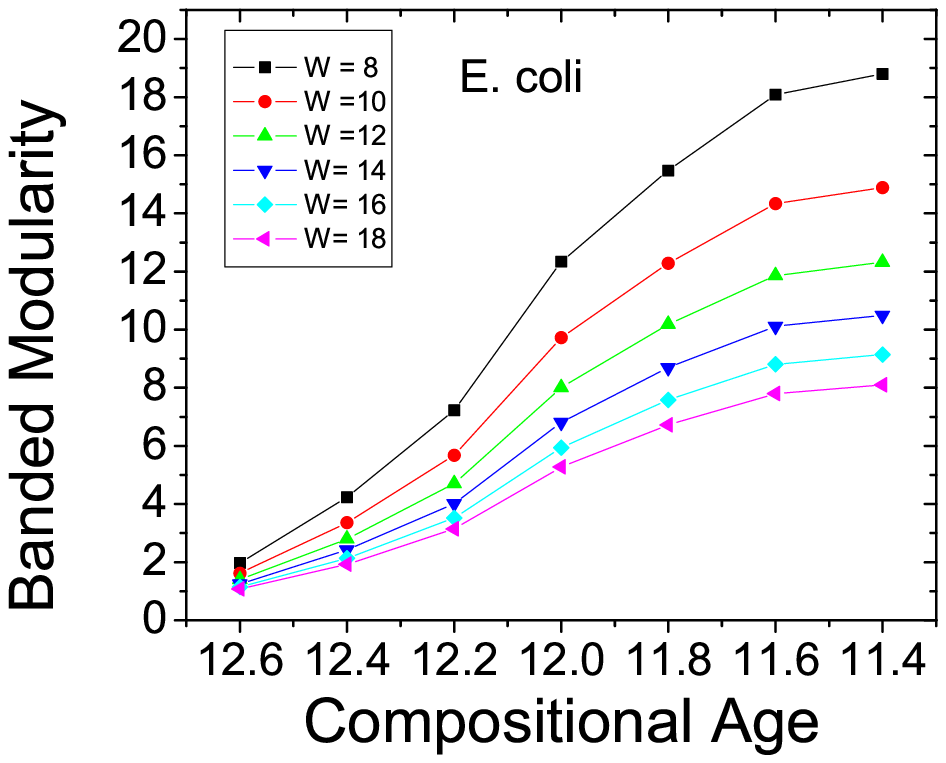}
\includegraphics[width=.4\textwidth]{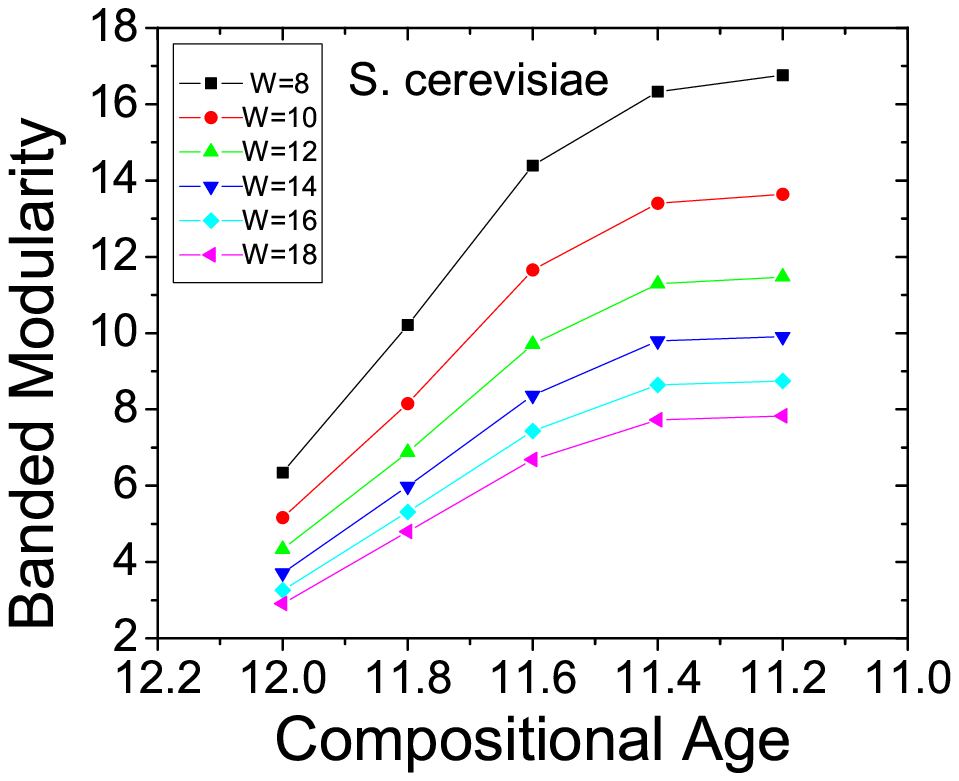}\\
\hspace{0.5in} (a) \hspace{2.2in} (b) \\
\end{center}
\caption[]{(color online) Evolution of banded modularity of 
the domain-domain interaction network in \emph{E.\ coli} (a) and
\emph{S.\ cerevisiae} (b). 
 \label{ddi_band}
}
\end{figure}

We observed modularity not only in the protein-protein interaction
network, but also in the domain-domain interaction
network.  We show the result of the banded modularity of the domain-domain
interaction network of \emph{E.\ coli} and
\emph{S.\ cerevisiae} in Fig.\ \ref{ddi_band}. The growth of banded modularity
is pronounced in both cases.

Our definitions of modularity allows the comparison of
modularity of matrices of different sizes. The saturated interaction matrix
does not have any modular structure, regardless of the band size,
as shown in Fig.\ \ref{ecoli_m}(e),(f).
A network generated by randomly selected proteins in \emph{E.\
coli} is of constant low modularity (see Appendix \ref{appendix2}),
independent of the number of proteins used.
The network
constructed based on its compositional age, however,
shows a clear growth of its modularity.
This result shows that the validity of organizing proteins by their
compositional age.

We also measure modularity of the unweighted domain-domain interaction network directly, without
construction of topological overlap matrix. 
We
determine the fraction of a protein to which other proteins interact.
To the extent that interactions become more localized within proteins, the
protein is defined to be more modular.
If protein B interacts with protein A, and the interaction is with only a few
of the domains of protein A, then this interaction is more modular than if protein B interacts
with a greater number of the domains of protein A. Averaging this measurement over all
proteins B, this procedure gives us a measure of the modularity of protein A.
So, we calculate the ratio of
interacting domains to the number of domains in a protein, which gives
the inverse of modularity. We define a
Score, which is the inverse of modularity, as:
\begin{equation}
Score: \quad \frac{1}{2N}\sum_{l=1}^{N} \left(\frac{I^A_l}{D^A_l
L^{2/3}_B}+\frac{I^B_l}{D^B_l L^{2/3}_A}\right)
\end{equation}
 Here $l$ represents a
protein-protein interaction or a link. To distinguish the two
proteins in a link, we mark one protein as A, the other one as B.
The number of links is $N$. The term $L_A$ ($L_B$) is the number of
amino acids of protein $A$ ($B$). The number of interacting
domains is $I_A$, and the number of total domains is $D^A$ in protein $A$. We
normalize the ratio of ${I^A}/{D_A}$ by the surface area of the
target protein $L_B^{2/3}$, and so the Score should measure only the
modularity and normalize out the size effect of target
proteins. 

\begin{figure}
\begin{center}
\includegraphics[width=.4\textwidth]{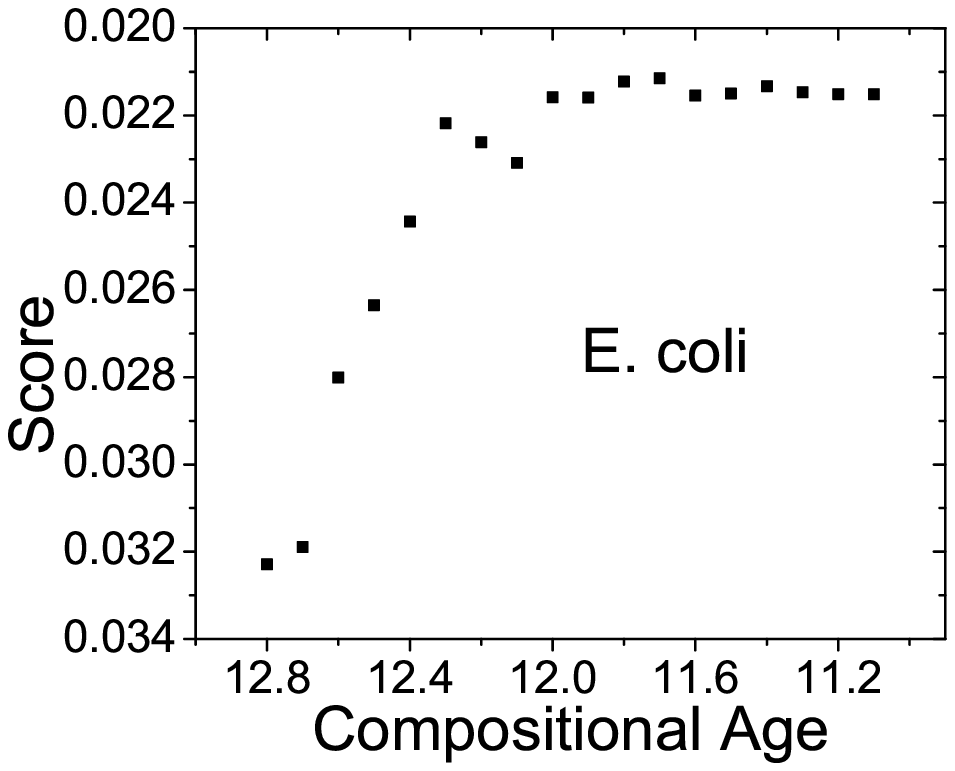}
\includegraphics[width=.4\textwidth]{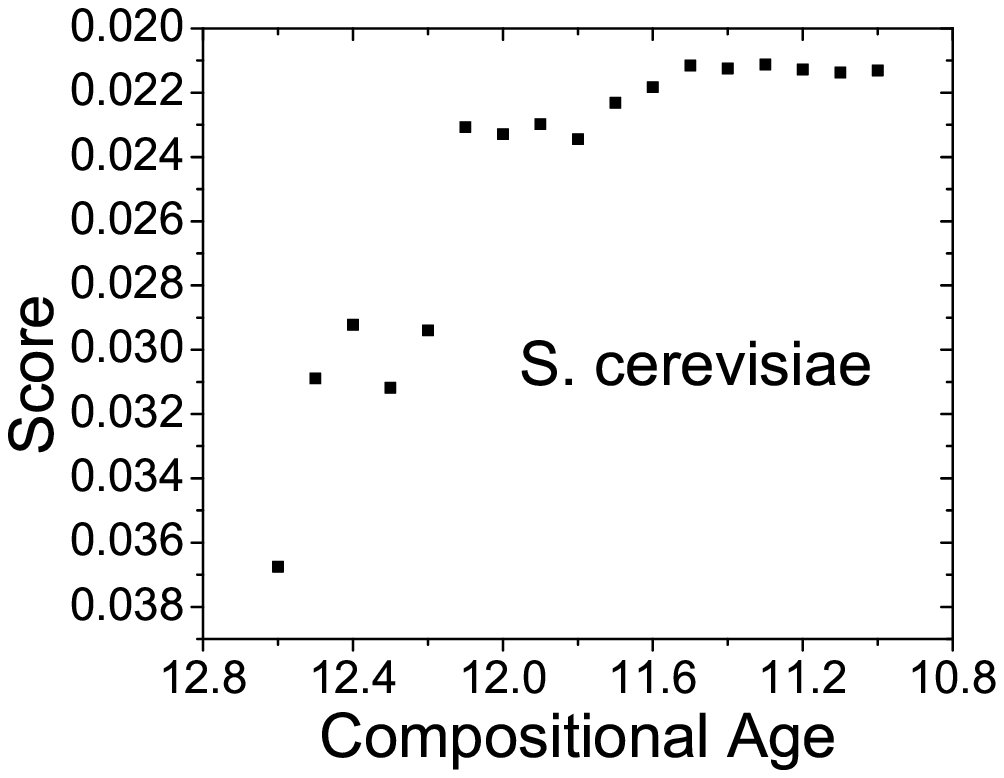}\\
\hspace{0.5in} (a) \hspace{2.2in} (b) \\
\end{center}
\caption[]{Domain interaction network modularity evolution in \emph{E.\ coli}
(a) and \emph{S.\ cerevisiae} (b).  Score is the inverse of modularity. 
 \label{ddi_score}
}
\end{figure}

In Fig.\ \ref{ddi_score}, we compare the Scores of different
domain-domain interaction network at different compositional age. 
The inverse of the Score increases monotonically with
evolutionary progress.
Because the inverse of the Score is modularity,
we again observe that 
modularity has increased through time.
This observation is robust under different definitions of the
Score (see Appendix \ref{appendix1}).

\subsubsection{Summary}
We have introduced several quantitative definitions of modularity for
interacting networks. We use them to measure the modularity of the
protein-protein interaction network
and domain-domain interaction network in \emph{S.\ cerevisiae} and 
\emph{E.\ coli}.
We have also introduced a method to quantify the evolutionary divergence
time of proteins.
 We consistently find
that
modularity, by all definitions and in both organisms, 
appears to have grown through time.
This observation is in agreement with the theory
that environmental change coupled with horizontal gene transfer naturally and
inevitably leads to evolution of increased modularity \cite{jun}.
In this sense, early life was a generalist, being less modular.
As evolution proceeded, and diversity of species increased and the environment
changed, proteins became more modular and specialized in their interactions.

\section{Conclusion}
The model results were described at the individual level.
In particular, we have presented the dynamics as that of individual
short protein sequences in a population.    The spin glass Hamiltonian,
however, is a general description for the replication rate in evolution.  
The spin glass Hamiltonian captures two basic features of evolution: 
evolution is relatively slow, and there are many local fitness optima.
Since the Hamiltonian captures the generic, basic features of evolution,
we expect the emergence of modularity to be a generic, fundamental
result.

Why is modularity so prevalent in the natural world?  Our hypothesis is
that a changing environment selects for
adaptable frameworks, and competition among different evolutionary frameworks
leads to selection of structures with the most efficient dynamics,
which are the modular ones.  
We have provided experimental evidence supporting this hypothesis.
We suggest that the beautiful, intricate, and interrelated structures
observed in nature may be the generic result of 
evolution in a changing environment.
The existence of such structure need not necessarily
rest on intelligent design or the anthropic principle.

It is now believed that large scale
exchange of genetic information is essential to increase the
rate of evolution \cite{Shapiro3,Colegrave,Goldenfeld2007}. 
Further experimental study of the relation between
large scale genetic exchange and the promotion of modularity
is warranted \cite{Lenski}. 
Some species of yeast may undergo either sexual or asexual
reproduction, and experiments suggest that yeasts undergoing
sexual reproduction are more evolvable \cite{Burt}. It would
be interesting to construct protocols to study the relation between sexual
recombination and modularity, possibly in gene expression
networks \cite{Bonhoeffer} in bacteria, in the laboratory.
At an applied level, we note that
the process by which antibiotics resistance evolved \cite{Walsh}
makes use of the modular structure of the genes encoding
the enzymes that degrade and the pumps that excrete antibiotics
and the modular structure of the proteins to which
antibiotics bind \cite{Maiden}.

\clearpage

\appendix

\section{Other Definitions of Domain Modularity}
\label{appendix1}

\begin{figure}
\begin{center}
\epsfig{file=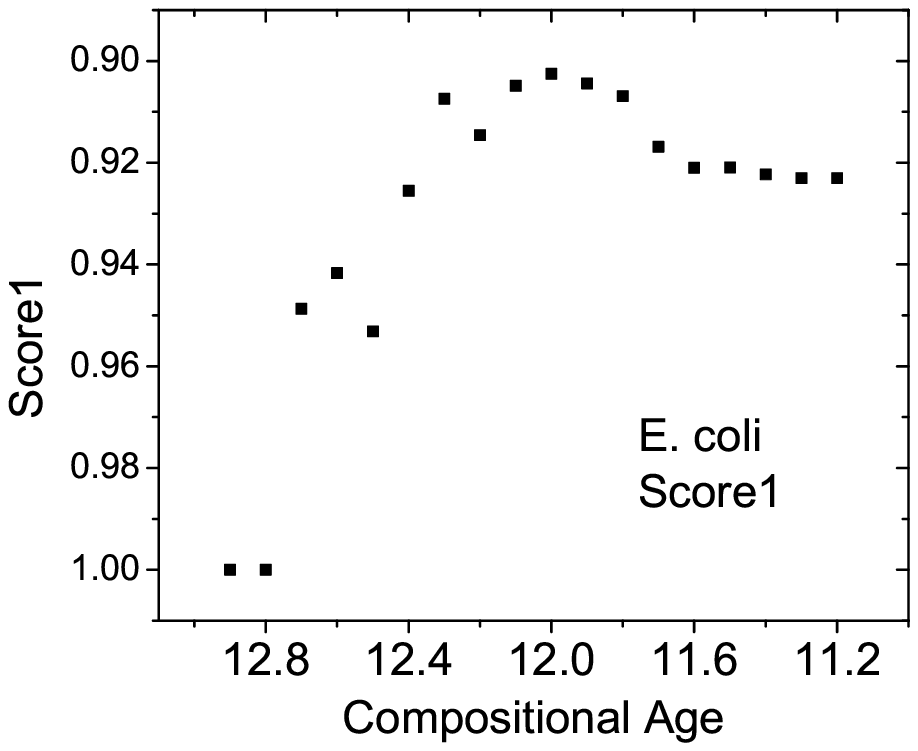,height=2.5in,clip=}
\epsfig{file=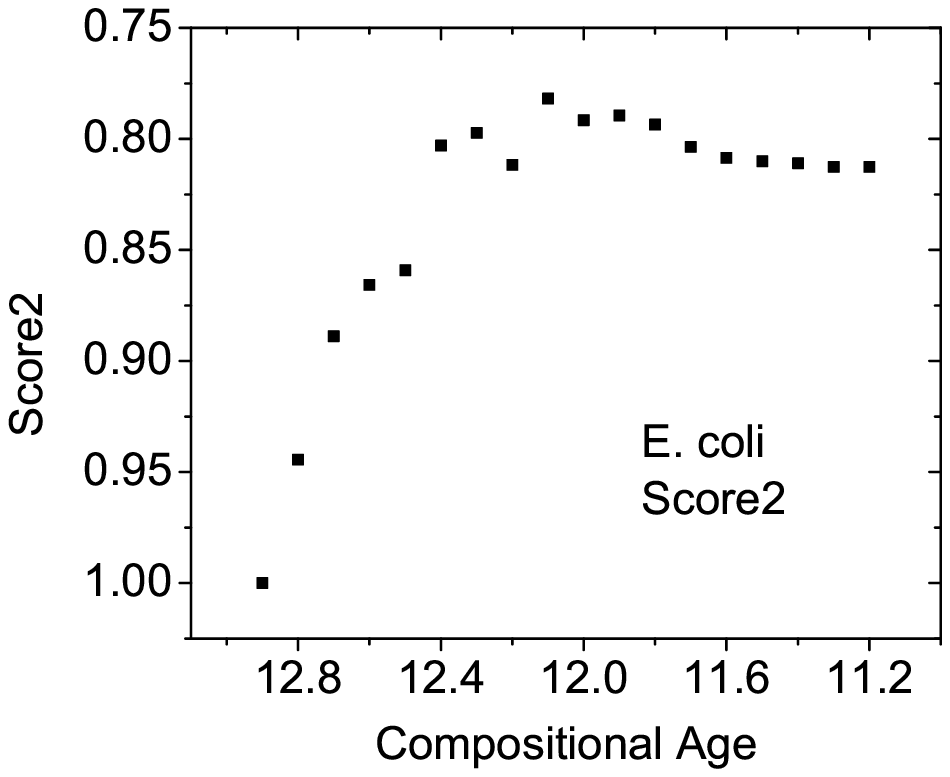,height=2.5in,clip=}\\
\epsfig{file=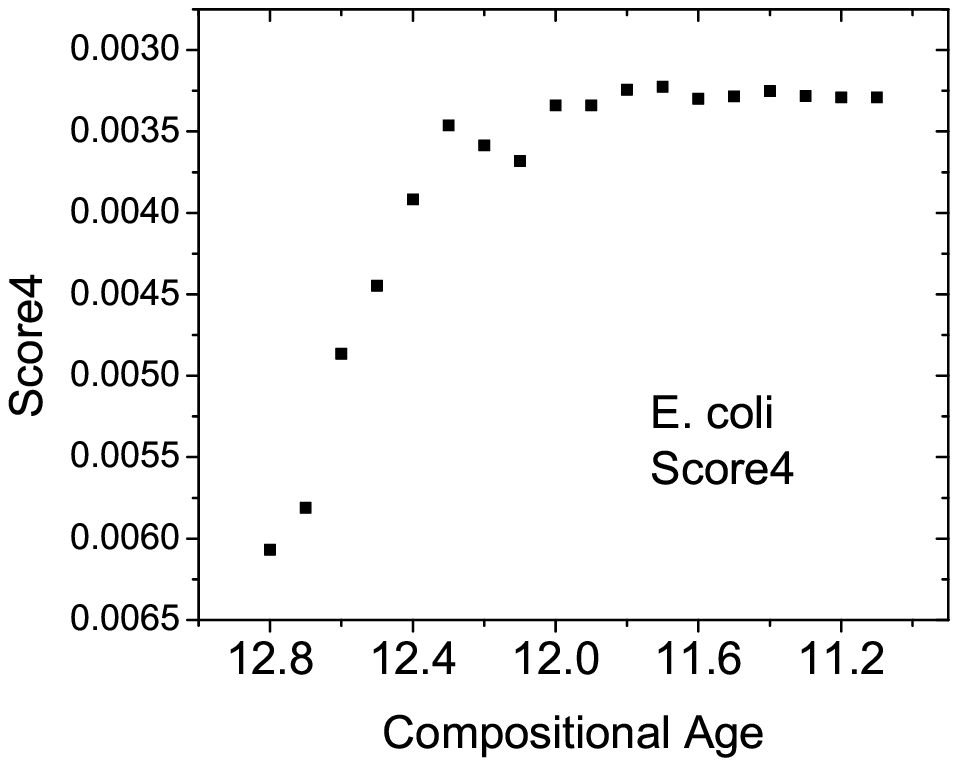,height=2.5in,clip=}
\epsfig{file=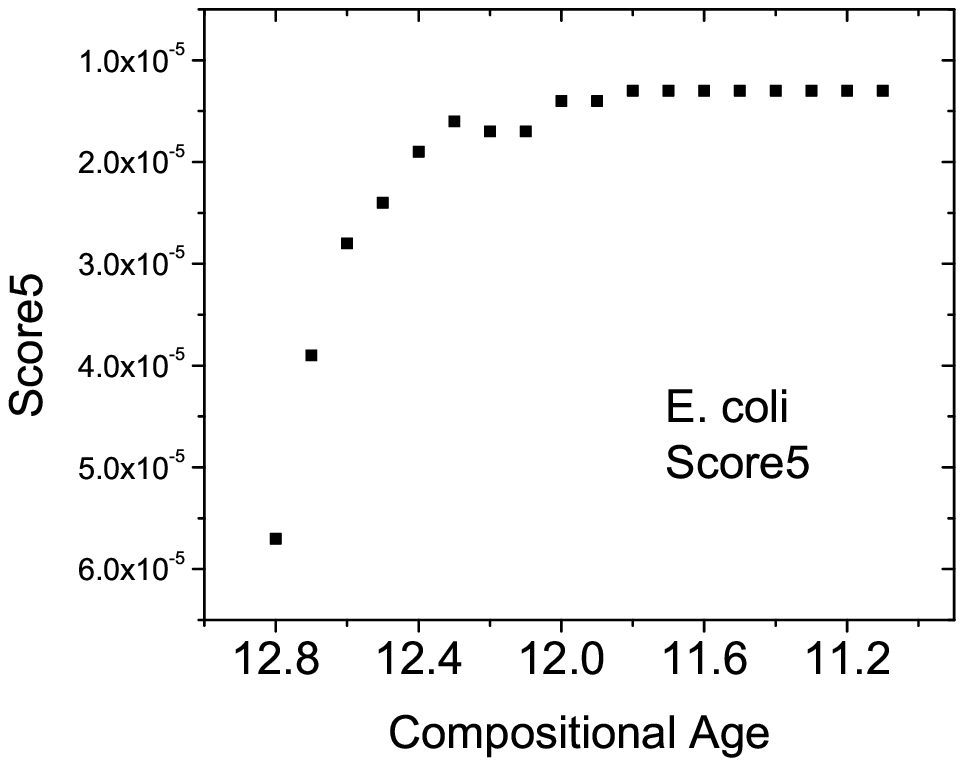,height=2.5in,clip=}\\
\epsfig{file=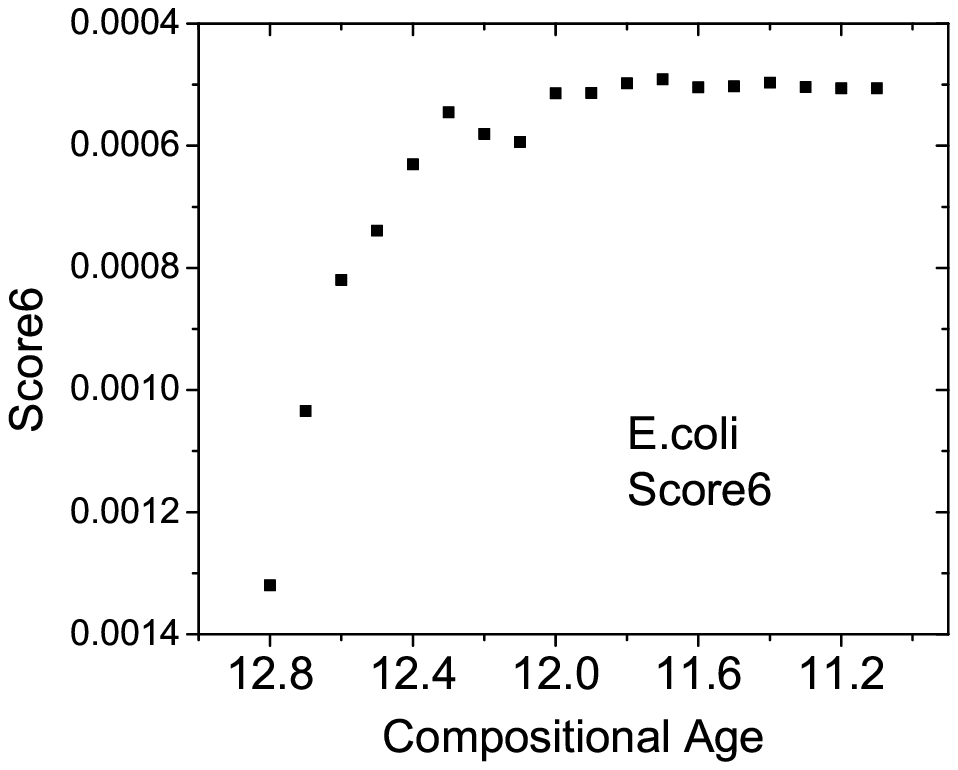,height=2.5in,clip=}
\epsfig{file=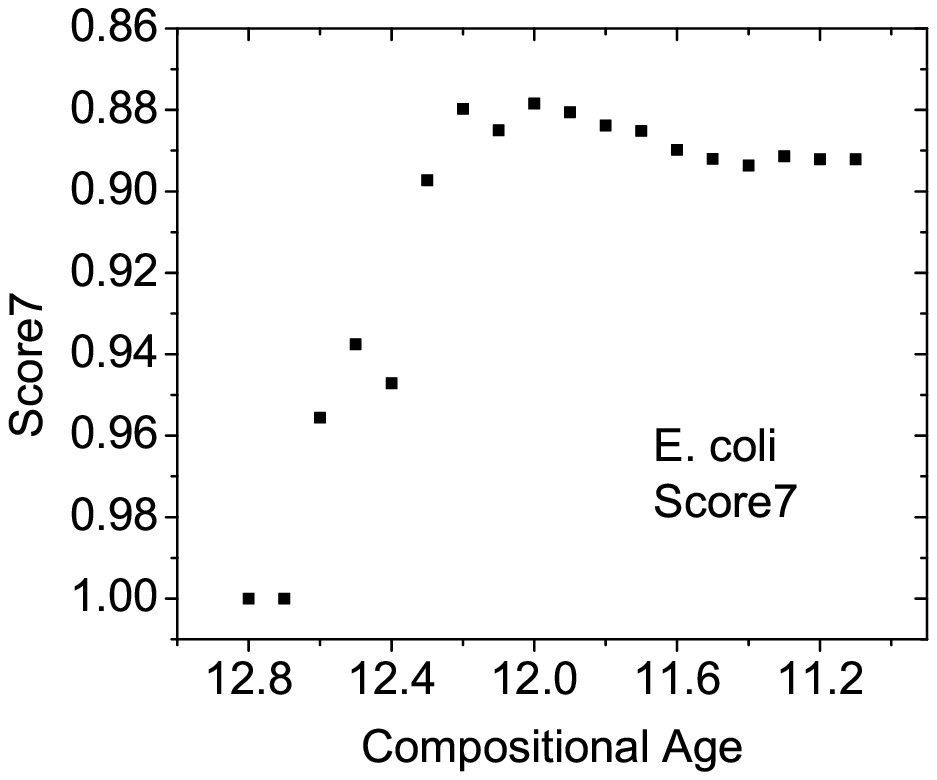,height=2.5in,clip=}\\
\end{center}
\caption{Different definitions of inverse modularity, 
for \emph{E.\ coli} and \emph{S.\ cerevisiae}.
\label{score7}
}
\end{figure}

We consider several different definitions of  a
measure of modularity in protein domain interactions:
\begin{eqnarray}
Score1: \quad  && \frac{1}{2N}\sum_{l=1}^{N}(\frac{I^A_l}{D^A_l
}+\frac{I^B_l}{D^B_l}) \\
Score2: \quad  &&\frac{1}{N}\sum_{l=1}^{N}(\frac{P_l}{D^A_l
D^B_l})\\
Score3:\quad  && \frac{1}{2N}\sum_{l=1}^{N}(\frac{I^A_l}{D^A_l
L^{2/3}_B}+\frac{I^B_l}{D^B_l L^{2/3}_A}) \\
Score4:\quad   &&\frac{1}{2N}\sum_{l=1}^{N}(\frac{I^A_l}{D^A_l
L^B}+\frac{I^B_l}{D^B_l L^A})\\
Score5:\quad   &&\frac{1}{N}\sum_{l=1}^{N}(\frac{P_l}{D^A_l
D^B_l L^A_l L^B_l})\\
Score6:\quad  && \frac{1}{N}\sum_{l=1}^{N}(\frac{P_l}{D^A_l
D^B_l (L^A_l)^{2/3} (L^B_l)^{2/3}})\\
Score7:\quad   &&\frac{1}{M}\sum_{l=1}^{N}(\sum_{j=1}^{D_B}\frac{T^A_j}{D^A_l
}+\sum_{j=1}^{D_A}\frac{T^B_j}{D^B_l})\\
\nonumber
\end{eqnarray}
In Score1--Score6, $l$ represents a protein-protein interaction link. To
distinguish the two proteins in each interaction link, we mark one
protein as A, and the other one B. The number of protein-protein
interaction links is $N$. 
The number of amino acids of protein $A$ ($B$)
is $L^A$ ($L^B$).
The number of total domains of protein $A$
is $D^A$.
The number of domain-domain interaction links
in the protein-protein interaction $l$
is $P_l$.
Score1 measures the
fraction of interacting domains to the total domains. Score2 
measures the saturation of domain interactions. Score3 is refinement
of Score1, excluding the effect of protein size by
normalizing with the surface area of the substrate protein. Score4
is another alternative, in which the fraction of available contacts
in the substrate is normalized simply by the
number of amino acids. Score5 and Score6 are advanced versions
of Score2, with normalizations for size of substrate. In Score7,  we
average over domain numbers instead of protein numbers. That is, when A
interacts with B, B has $D_B$ domains; for the $j$th domain in
protein B, it can interact with $T^A_j$ domains in protein A, and
$M=\sum_{l=1}^{N}(D_B+D_A)$. Score1--Score7 are all measures of the
inverse of modularity.  All of these scores show an increase of
modularity through time,
 see Fig.\ \ref{score7}. 

\begin{figure}
\begin{center}
\epsfig{file=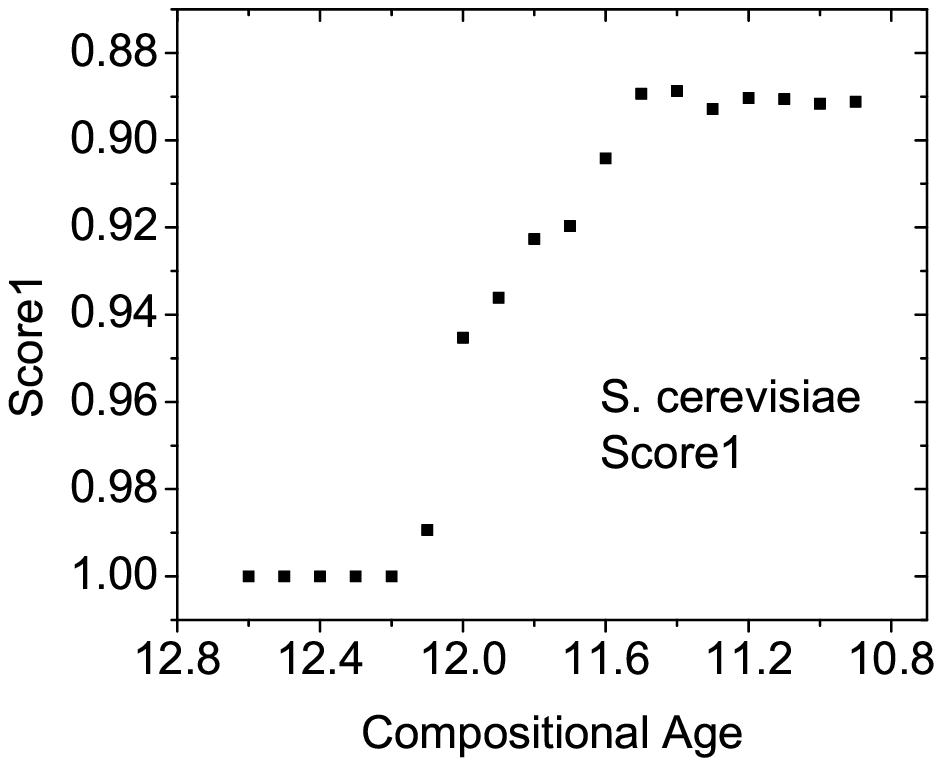,height=2.5in,clip=}
\epsfig{file=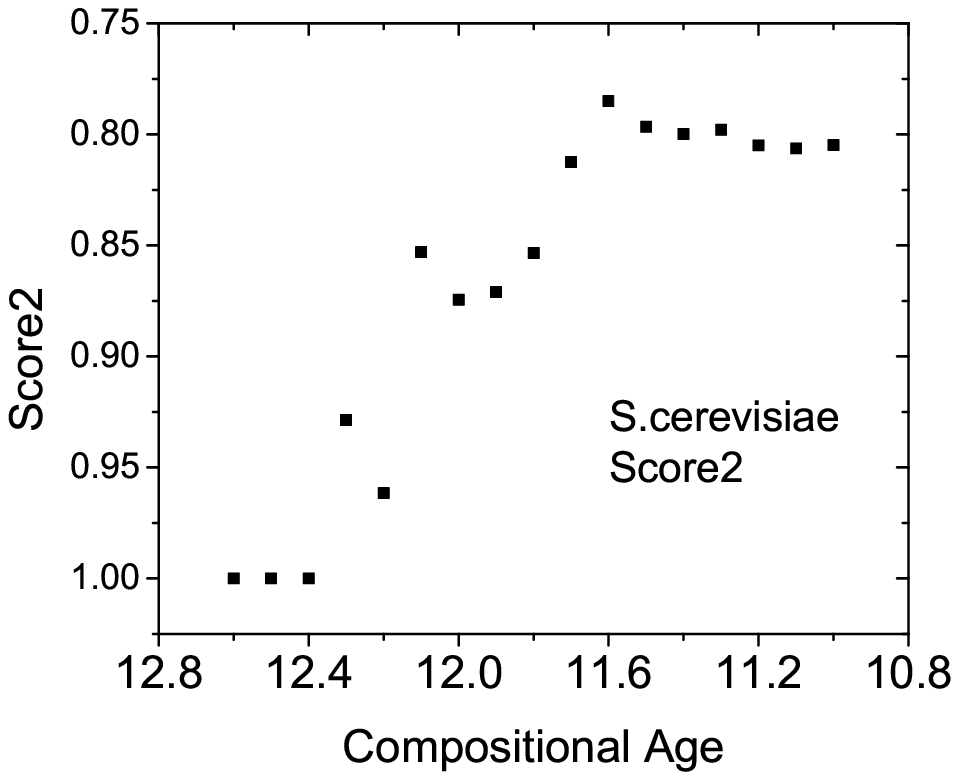,height=2.5in,clip=}\\
\epsfig{file=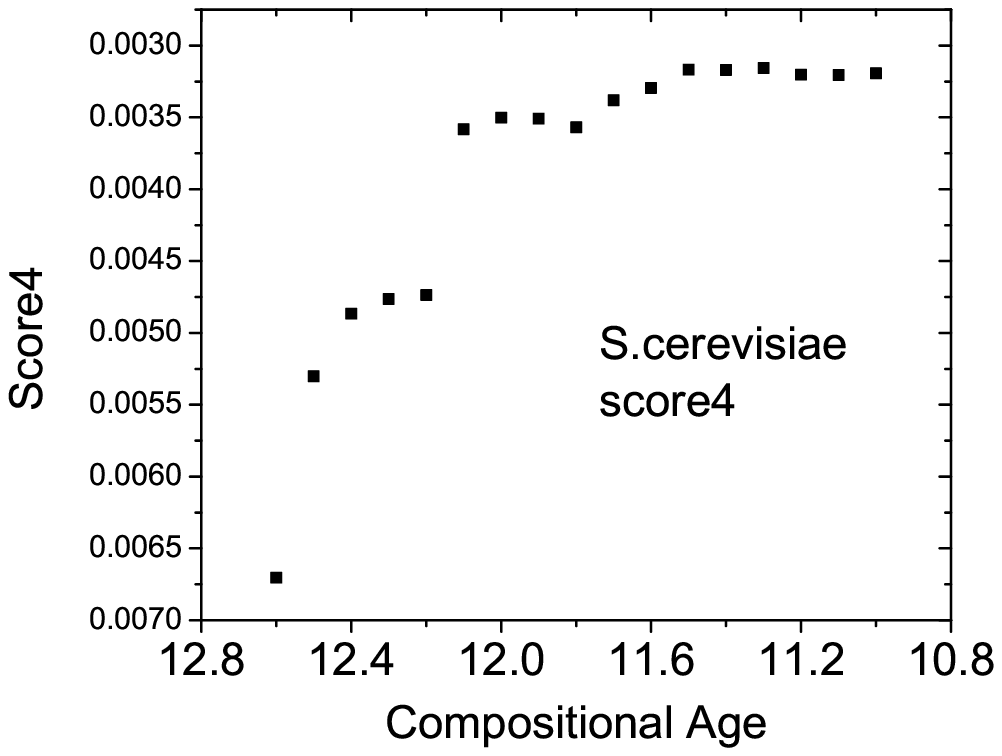,height=2.5in,clip=}
\epsfig{file=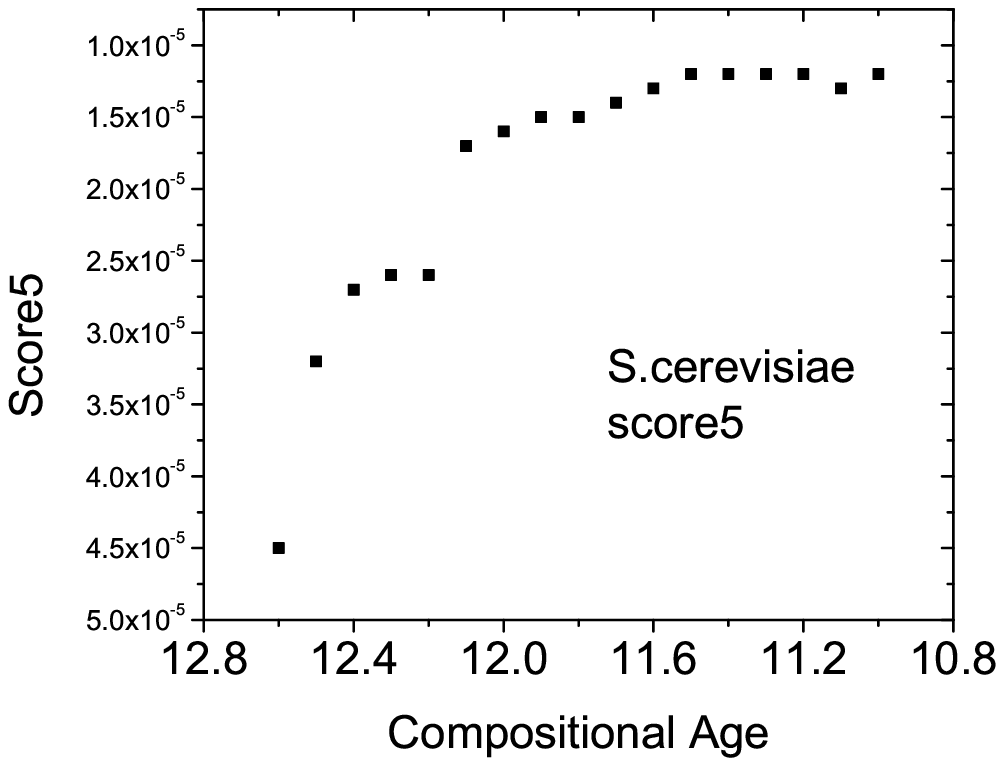,height=2.5in,clip=}\\
\epsfig{file=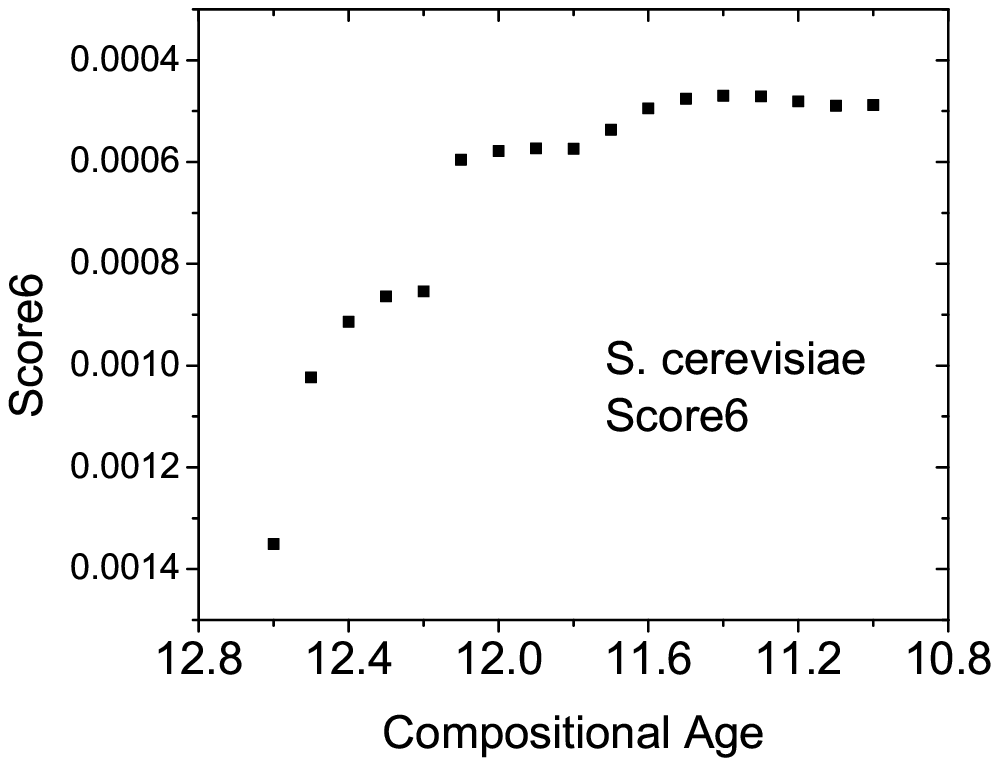,height=2.5in,clip=}
\epsfig{file=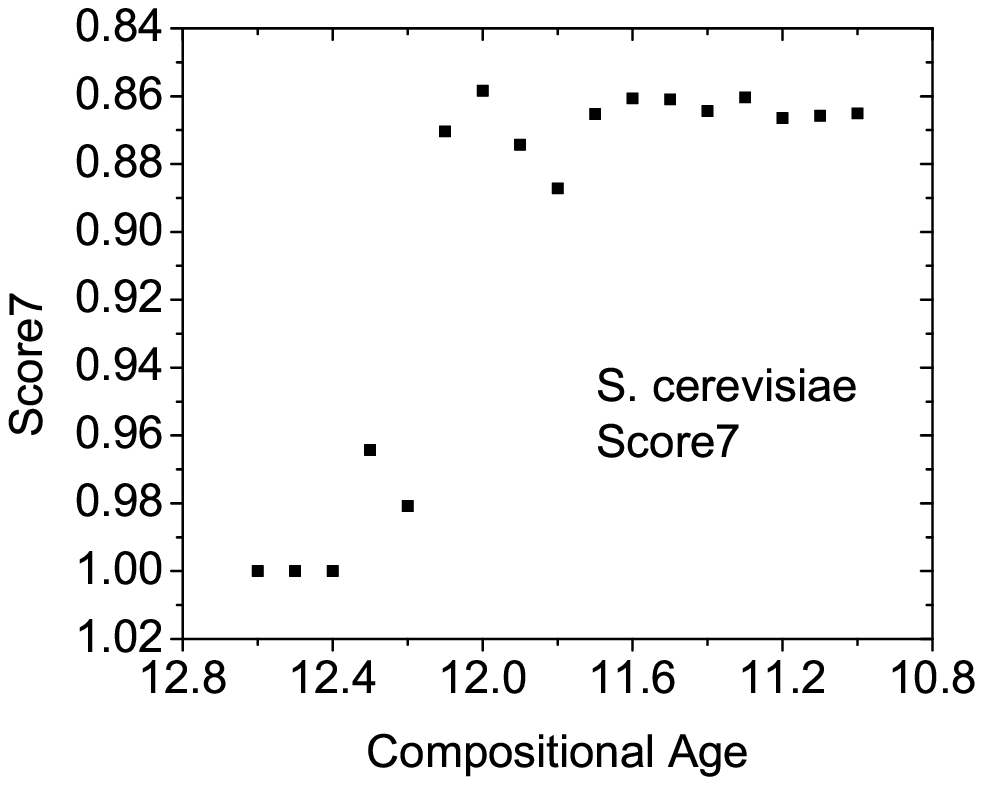,height=2.5in,clip=}\\
\end{center}
\label{scorecere}
\end{figure}

\section{Random Networks are Not Modular}
\label{appendix2}
\begin{figure}
\begin{center}
\epsfig{file=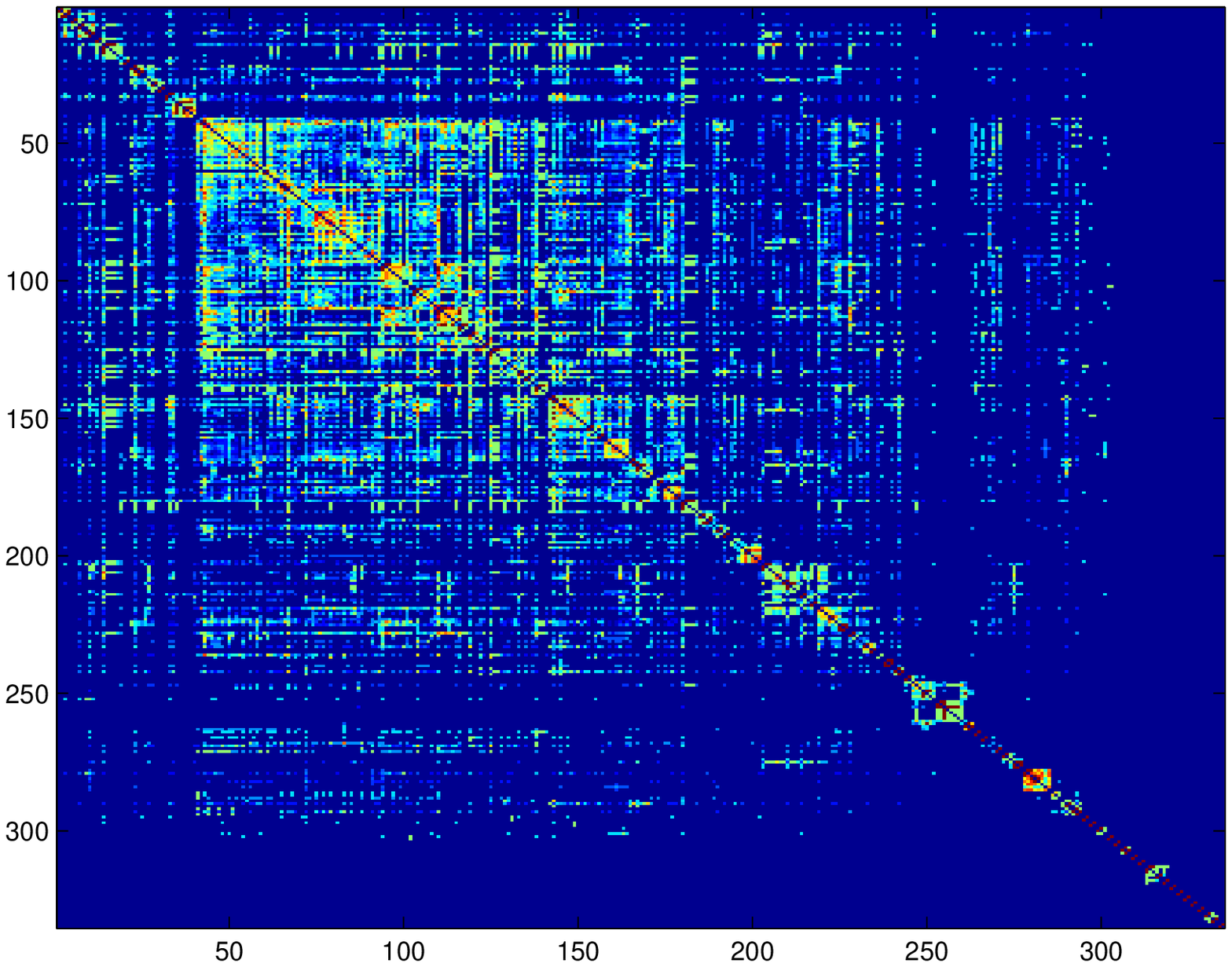,height=4in,clip=}
\epsfig{file=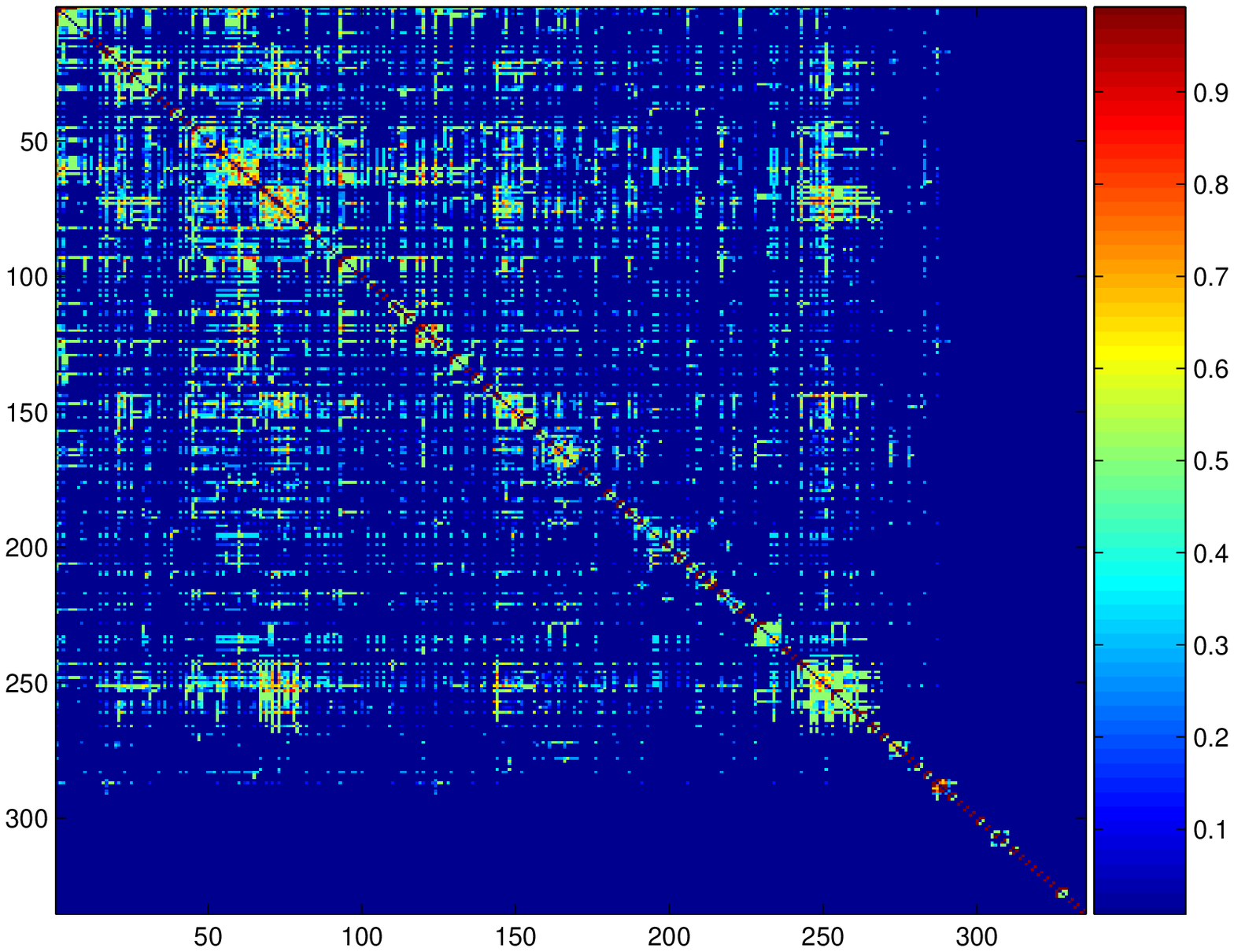,height=4in,clip=}\\
\hspace{0.3in} (a) \hspace{2.5in}(b) \\
\end{center}
\caption{(color online) (a) A random network. (b) \emph{E.\ coli} protein interaction
network  at compositional age 12.2.
 \label{random}
}
\end{figure}

\begin{figure}
\begin{center}
\epsfig{file=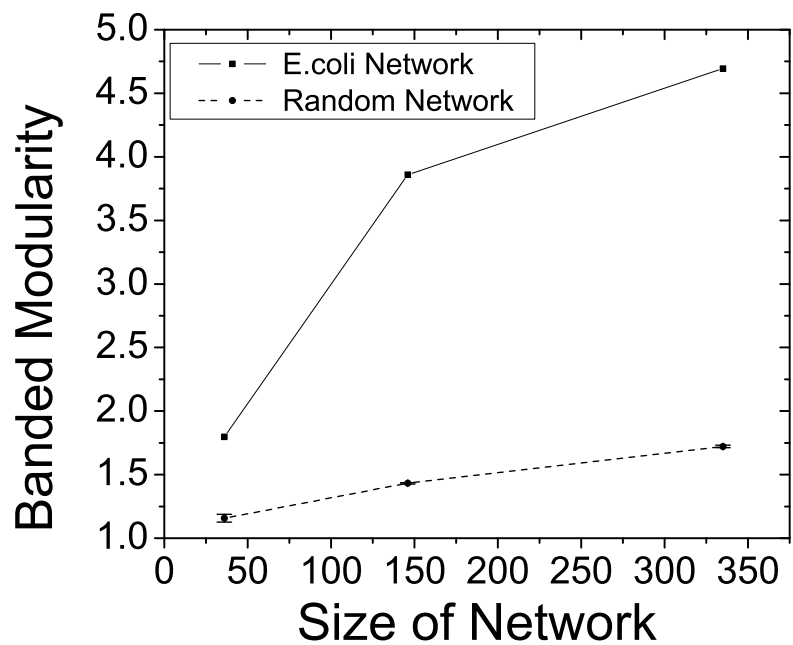,height=4in,clip=}\\
\end{center}
\caption{Comparison of banded modularity with width 
8 between the \emph{E.\ coli} network and the random network. 
The red line is the banded modularity of random network. The
black line is the \emph{E.\ coli}  network with size 36 
at compositional age 12.6, size
335 at compositional age 12.2, 
and size 949 at compositional age 11.8. 
Error bars are shown in blue.
 \label{size}
}
\end{figure}

We select 352 proteins in \emph{E.\ coli} at random and find the
interaction in DIP, then we construct the interaction network.
The result 
after clustering, for \emph{E.\ coli} at compositional age 12.2, is
shown in 
Fig.\ \ref{random}. The \emph{E.\ coli} network shows
hierarchical structure, while the random network has no hierarchical
structure. The selection by compositional age elucidates the
non-random effects of evolution.
We also use the random network to test the quality of our
definition of block modularity. First, we calculate 
the degree of \emph{E.\ coli}
protein interaction network at different compositional ages, then, we construct
several random networks with the same size and degree as the
\emph{E.\ coli} networks so constructed.
We repeat this procedure 10 times for each point.  We use
the average linkage hierarchical clustering method to calculate the
block modularity. We make a comparison with the \emph{E.\ coli} network
selected based on compositional age in Fig.\ \ref{size}. 
The \emph{E.\ coli} networks are much more
modular than are the random networks; the
modularity of random networks is due only to random fluctuations
that are grouped by the hierarchical clustering algorithm.
The modules are visually apparent in the clustered matrix, as in
Fig.\ \ref{random}(b).  

\clearpage


\bibliography{modularity}

\begin{thebibliography}{49}
\expandafter\ifx\csname natexlab\endcsname\relax\def\natexlab#1{#1}\fi
\expandafter\ifx\csname bibnamefont\endcsname\relax
  \def\bibnamefont#1{#1}\fi
\expandafter\ifx\csname bibfnamefont\endcsname\relax
  \def\bibfnamefont#1{#1}\fi
\expandafter\ifx\csname citenamefont\endcsname\relax
  \def\citenamefont#1{#1}\fi
\expandafter\ifx\csname url\endcsname\relax
  \def\url#1{\texttt{#1}}\fi
\expandafter\ifx\csname urlprefix\endcsname\relax\def\urlprefix{URL }\fi
\providecommand{\bibinfo}[2]{#2}
\providecommand{\eprint}[2][]{\url{#2}}

\bibitem[{\citenamefont{Shapiro}(2004)}]{Shapiro2004}
\bibinfo{author}{\bibfnamefont{J.~A.} \bibnamefont{Shapiro}},
  \bibinfo{journal}{Gene} \textbf{\bibinfo{volume}{345}}, \bibinfo{pages}{91}
  (\bibinfo{year}{2004}).

\bibitem[{\citenamefont{Shapiro}(2005)}]{Shapiro2005}
\bibinfo{author}{\bibfnamefont{J.~A.} \bibnamefont{Shapiro}},
  \bibinfo{journal}{BioEssays} \textbf{\bibinfo{volume}{27}},
  \bibinfo{pages}{122} (\bibinfo{year}{2005}).

\bibitem[{\citenamefont{Misevic et~al.}(2006)\citenamefont{Misevic, Ofria, and
  Lenski}}]{Lenski}
\bibinfo{author}{\bibfnamefont{D.}~\bibnamefont{Misevic}},
  \bibinfo{author}{\bibfnamefont{C.}~\bibnamefont{Ofria}}, \bibnamefont{and}
  \bibinfo{author}{\bibfnamefont{R.~E.} \bibnamefont{Lenski}},
  \bibinfo{journal}{Proc. R. Soc. B} \textbf{\bibinfo{volume}{273}},
  \bibinfo{pages}{457} (\bibinfo{year}{2006}).

\bibitem[{\citenamefont{Shapiro}(1992)}]{Shapiro}
\bibinfo{author}{\bibfnamefont{J.~A.} \bibnamefont{Shapiro}},
  \bibinfo{journal}{Genetica} \textbf{\bibinfo{volume}{86}},
  \bibinfo{pages}{99} (\bibinfo{year}{1992}).

\bibitem[{\citenamefont{Goldenfeld and Woese}(2007)}]{Goldenfeld2007}
\bibinfo{author}{\bibfnamefont{N.}~\bibnamefont{Goldenfeld}} \bibnamefont{and}
  \bibinfo{author}{\bibfnamefont{C.}~\bibnamefont{Woese}},
  \bibinfo{journal}{Nature} \textbf{\bibinfo{volume}{445}},
  \bibinfo{pages}{369} (\bibinfo{year}{2007}).

\bibitem[{\citenamefont{Lipson et~al.}(2002)\citenamefont{Lipson, Pollack, and
  Suh}}]{Lipson2002}
\bibinfo{author}{\bibfnamefont{H.}~\bibnamefont{Lipson}},
  \bibinfo{author}{\bibfnamefont{J.~B.} \bibnamefont{Pollack}},
  \bibnamefont{and} \bibinfo{author}{\bibfnamefont{N.~P.} \bibnamefont{Suh}},
  \bibinfo{journal}{Evolution} \textbf{\bibinfo{volume}{56}},
  \bibinfo{pages}{1549} (\bibinfo{year}{2002}).

\bibitem[{\citenamefont{Tanase-Nicola et~al.}(2006)\citenamefont{Tanase-Nicola,
  Warren, and tenWolde}}]{Wolde}
\bibinfo{author}{\bibfnamefont{S.}~\bibnamefont{Tanase-Nicola}},
  \bibinfo{author}{\bibfnamefont{P.~B.} \bibnamefont{Warren}},
  \bibnamefont{and} \bibinfo{author}{\bibfnamefont{P.~R.}
  \bibnamefont{tenWolde}}, \bibinfo{journal}{Phys. Rev. Lett.}
  \textbf{\bibinfo{volume}{97}}, \bibinfo{pages}{068102}
  (\bibinfo{year}{2006}).

\bibitem[{\citenamefont{Variano et~al.}(2004)\citenamefont{Variano, McCoy, and
  Lipson}}]{Lipson2004}
\bibinfo{author}{\bibfnamefont{E.~A.} \bibnamefont{Variano}},
  \bibinfo{author}{\bibfnamefont{J.~H.} \bibnamefont{McCoy}}, \bibnamefont{and}
  \bibinfo{author}{\bibfnamefont{H.}~\bibnamefont{Lipson}},
  \bibinfo{journal}{Phys. Rev. Lett.} \textbf{\bibinfo{volume}{92}},
  \bibinfo{pages}{188701} (\bibinfo{year}{2004}).

\bibitem[{\citenamefont{Shen-Orr et~al.}(2002)\citenamefont{Shen-Orr, Milo,
  Mangan, and Alon}}]{Alon2002}
\bibinfo{author}{\bibfnamefont{S.~S.} \bibnamefont{Shen-Orr}},
  \bibinfo{author}{\bibfnamefont{R.}~\bibnamefont{Milo}},
  \bibinfo{author}{\bibfnamefont{S.}~\bibnamefont{Mangan}}, \bibnamefont{and}
  \bibinfo{author}{\bibfnamefont{U.}~\bibnamefont{Alon}},
  \bibinfo{journal}{Nature Genetics} \textbf{\bibinfo{volume}{31}},
  \bibinfo{pages}{64} (\bibinfo{year}{2002}).

\bibitem[{\citenamefont{Kashtan and Alon}(2005)}]{Alon2005}
\bibinfo{author}{\bibfnamefont{N.}~\bibnamefont{Kashtan}} \bibnamefont{and}
  \bibinfo{author}{\bibfnamefont{U.}~\bibnamefont{Alon}},
  \bibinfo{journal}{Proc. Natl. Acad. Sci. USA} \textbf{\bibinfo{volume}{102}},
  \bibinfo{pages}{13773} (\bibinfo{year}{2005}).

\bibitem[{\citenamefont{Vetsigian et~al.}(2006)\citenamefont{Vetsigian, Woese,
  and Goldenfeld}}]{Goldenfeld2006}
\bibinfo{author}{\bibfnamefont{K.}~\bibnamefont{Vetsigian}},
  \bibinfo{author}{\bibfnamefont{C.}~\bibnamefont{Woese}}, \bibnamefont{and}
  \bibinfo{author}{\bibfnamefont{N.}~\bibnamefont{Goldenfeld}},
  \bibinfo{journal}{Proc. Natl. Acad. Sci. USA} \textbf{\bibinfo{volume}{103}},
  \bibinfo{pages}{10696} (\bibinfo{year}{2006}).

\bibitem[{\citenamefont{Csete and Doyle}(2002)}]{Doyle}
\bibinfo{author}{\bibfnamefont{M.~E.} \bibnamefont{Csete}} \bibnamefont{and}
  \bibinfo{author}{\bibfnamefont{J.~C.} \bibnamefont{Doyle}},
  \bibinfo{journal}{Science} \textbf{\bibinfo{volume}{295}},
  \bibinfo{pages}{1664} (\bibinfo{year}{2002}).

\bibitem[{\citenamefont{Kitano}(2004)}]{Kitano}
\bibinfo{author}{\bibfnamefont{H.}~\bibnamefont{Kitano}},
  \bibinfo{journal}{Nature Reviews Genetics} \textbf{\bibinfo{volume}{5}},
  \bibinfo{pages}{826} (\bibinfo{year}{2004}).

\bibitem[{\citenamefont{Oikonomou and Cluzel}(2006)}]{Cluzel}
\bibinfo{author}{\bibfnamefont{P.}~\bibnamefont{Oikonomou}} \bibnamefont{and}
  \bibinfo{author}{\bibfnamefont{P.}~\bibnamefont{Cluzel}},
  \bibinfo{journal}{Nature Physics} \textbf{\bibinfo{volume}{2}},
  \bibinfo{pages}{532} (\bibinfo{year}{2006}).

\bibitem[{\citenamefont{Earl and Deem}(2004)}]{Earl}
\bibinfo{author}{\bibfnamefont{D.~J.} \bibnamefont{Earl}} \bibnamefont{and}
  \bibinfo{author}{\bibfnamefont{M.~W.} \bibnamefont{Deem}},
  \bibinfo{journal}{Proc. Natl. Acad. Sci. USA} \textbf{\bibinfo{volume}{101}},
  \bibinfo{pages}{11531} (\bibinfo{year}{2004}).

\bibitem[{\citenamefont{Deem}(2007)}]{Deem2007}
\bibinfo{author}{\bibfnamefont{M.~W.} \bibnamefont{Deem}},
  \bibinfo{journal}{Physics Today} \textbf{\bibinfo{volume}{60}},
  \bibinfo{pages}{42} (\bibinfo{year}{2007}).

\bibitem[{\citenamefont{Gardener and Zuidema}(2003)}]{Gardner2003}
\bibinfo{author}{\bibfnamefont{A.}~\bibnamefont{Gardener}} \bibnamefont{and}
  \bibinfo{author}{\bibfnamefont{W.}~\bibnamefont{Zuidema}},
  \bibinfo{journal}{Evolution} \textbf{\bibinfo{volume}{57}},
  \bibinfo{pages}{1448} (\bibinfo{year}{2003}).

\bibitem[{\citenamefont{Sun and Deem}(2007)}]{jun}
\bibinfo{author}{\bibfnamefont{J.}~\bibnamefont{Sun}} \bibnamefont{and}
  \bibinfo{author}{\bibfnamefont{M.~W.} \bibnamefont{Deem}},
  \bibinfo{journal}{Phys. Rev. Lett.} \textbf{\bibinfo{volume}{99}},
  \bibinfo{pages}{228107} (\bibinfo{year}{2007}).

\bibitem[{\citenamefont{Kauffman and Levin}(1987)}]{Kauffman}
\bibinfo{author}{\bibfnamefont{S.}~\bibnamefont{Kauffman}} \bibnamefont{and}
  \bibinfo{author}{\bibfnamefont{S.}~\bibnamefont{Levin}}, \bibinfo{journal}{J.
  Theor. Biol.} \textbf{\bibinfo{volume}{128}}, \bibinfo{pages}{11}
  (\bibinfo{year}{1987}).

\bibitem[{\citenamefont{Deem and Lee}(2003)}]{Deem}
\bibinfo{author}{\bibfnamefont{M.~W.} \bibnamefont{Deem}} \bibnamefont{and}
  \bibinfo{author}{\bibfnamefont{H.-Y.} \bibnamefont{Lee}},
  \bibinfo{journal}{Phys. Rev. Lett.} \textbf{\bibinfo{volume}{91}},
  \bibinfo{pages}{068101} (\bibinfo{year}{2003}).

\bibitem[{\citenamefont{Sun et~al.}(2005)\citenamefont{Sun, Earl, and
  Deem}}]{Sun}
\bibinfo{author}{\bibfnamefont{J.}~\bibnamefont{Sun}},
  \bibinfo{author}{\bibfnamefont{D.~J.} \bibnamefont{Earl}}, \bibnamefont{and}
  \bibinfo{author}{\bibfnamefont{M.~W.} \bibnamefont{Deem}},
  \bibinfo{journal}{Phys. Rev. Lett.} \textbf{\bibinfo{volume}{95}},
  \bibinfo{pages}{148104} (\bibinfo{year}{2005}).

\bibitem[{\citenamefont{Anderson}(1983)}]{Anderson}
\bibinfo{author}{\bibfnamefont{P.~W.} \bibnamefont{Anderson}},
  \bibinfo{journal}{Proc. Natl. Acad. Sci. USA} \textbf{\bibinfo{volume}{80}},
  \bibinfo{pages}{3386} (\bibinfo{year}{1983}).

\bibitem[{\citenamefont{Stein and Anderson}(1984)}]{Stein}
\bibinfo{author}{\bibfnamefont{D.~L.} \bibnamefont{Stein}} \bibnamefont{and}
  \bibinfo{author}{\bibfnamefont{P.~W.} \bibnamefont{Anderson}},
  \bibinfo{journal}{Proc. Natl. Acad. Sci. USA} \textbf{\bibinfo{volume}{81}},
  \bibinfo{pages}{1751} (\bibinfo{year}{1984}).

\bibitem[{\citenamefont{Perelson and Macken}(1995)}]{Perelson}
\bibinfo{author}{\bibfnamefont{A.~S.} \bibnamefont{Perelson}} \bibnamefont{and}
  \bibinfo{author}{\bibfnamefont{C.~A.} \bibnamefont{Macken}},
  \bibinfo{journal}{Proc. Natl. Acad. Sci. USA} \textbf{\bibinfo{volume}{92}},
  \bibinfo{pages}{9657} (\bibinfo{year}{1995}).

\bibitem[{\citenamefont{Mezard et~al.}(1987)\citenamefont{Mezard, Parisi, and
  Virasoro}}]{Mezard}
\bibinfo{author}{\bibfnamefont{M.}~\bibnamefont{Mezard}},
  \bibinfo{author}{\bibfnamefont{G.}~\bibnamefont{Parisi}}, \bibnamefont{and}
  \bibinfo{author}{\bibfnamefont{M.}~\bibnamefont{Virasoro}},
  \emph{\bibinfo{title}{Spin Glass Theory and Beyond: An Introduction to the
  Replica Method and Its Applications}} (\bibinfo{publisher}{World Scientific},
  \bibinfo{address}{New Jersey}, \bibinfo{year}{1987}).

\bibitem[{\citenamefont{Bogarad and Deem}(1999)}]{Bogarad}
\bibinfo{author}{\bibfnamefont{L.~D.} \bibnamefont{Bogarad}} \bibnamefont{and}
  \bibinfo{author}{\bibfnamefont{M.~W.} \bibnamefont{Deem}},
  \bibinfo{journal}{Proc. Natl. Acad. Sci. USA} \textbf{\bibinfo{volume}{96}},
  \bibinfo{pages}{2591} (\bibinfo{year}{1999}).

\bibitem[{\citenamefont{Sun et~al.}(2006)\citenamefont{Sun, Earl, and
  Deem}}]{Sun2006}
\bibinfo{author}{\bibfnamefont{J.}~\bibnamefont{Sun}},
  \bibinfo{author}{\bibfnamefont{D.~J.} \bibnamefont{Earl}}, \bibnamefont{and}
  \bibinfo{author}{\bibfnamefont{M.~W.} \bibnamefont{Deem}},
  \bibinfo{journal}{Mod. Phys. Lett. B} \textbf{\bibinfo{volume}{20}},
  \bibinfo{pages}{63} (\bibinfo{year}{2006}).

\bibitem[{\citenamefont{Park and Deem}(2007)}]{Park}
\bibinfo{author}{\bibfnamefont{J.-M.} \bibnamefont{Park}} \bibnamefont{and}
  \bibinfo{author}{\bibfnamefont{M.~W.} \bibnamefont{Deem}},
  \bibinfo{journal}{Phys. Rev. Lett.} \textbf{\bibinfo{volume}{98}},
  \bibinfo{pages}{058101} (\bibinfo{year}{2007}).

\bibitem[{\citenamefont{Barab\'{a}si and
  Oltvai}(2004)}]{barabasi_networkbiology}
\bibinfo{author}{\bibfnamefont{A.~L.} \bibnamefont{Barab\'{a}si}}
  \bibnamefont{and} \bibinfo{author}{\bibfnamefont{Z.~N.}
  \bibnamefont{Oltvai}}, \bibinfo{journal}{Nat. Rev. Genet.}
  \textbf{\bibinfo{volume}{5}}, \bibinfo{pages}{101} (\bibinfo{year}{2004}).

\bibitem[{\citenamefont{Gammaitoni et~al.}(1998)\citenamefont{Gammaitoni,
  H{\"a}nggi, Jung, and Marchesoni}}]{Hanngi1998}
\bibinfo{author}{\bibfnamefont{L.}~\bibnamefont{Gammaitoni}},
  \bibinfo{author}{\bibfnamefont{P.}~\bibnamefont{H{\"a}nggi}},
  \bibinfo{author}{\bibfnamefont{P.}~\bibnamefont{Jung}}, \bibnamefont{and}
  \bibinfo{author}{\bibfnamefont{F.}~\bibnamefont{Marchesoni}},
  \bibinfo{journal}{Rev. Mod. Phys.} \textbf{\bibinfo{volume}{70}},
  \bibinfo{pages}{223} (\bibinfo{year}{1998}).

\bibitem[{\citenamefont{Sato et~al.}(2003)\citenamefont{Sato, Ito, Yomo, and
  Kaneko}}]{Kaneko}
\bibinfo{author}{\bibfnamefont{K.}~\bibnamefont{Sato}},
  \bibinfo{author}{\bibfnamefont{Y.}~\bibnamefont{Ito}},
  \bibinfo{author}{\bibfnamefont{T.}~\bibnamefont{Yomo}}, \bibnamefont{and}
  \bibinfo{author}{\bibfnamefont{K.}~\bibnamefont{Kaneko}},
  \bibinfo{journal}{Proc. Natl. Acad. Sci. USA} \textbf{\bibinfo{volume}{100}},
  \bibinfo{pages}{14086} (\bibinfo{year}{2003}).

\bibitem[{\citenamefont{Parter et~al.}(2007)\citenamefont{Parter, Kashtan, and
  Alon}}]{parter2007}
\bibinfo{author}{\bibfnamefont{M.}~\bibnamefont{Parter}},
  \bibinfo{author}{\bibfnamefont{N.}~\bibnamefont{Kashtan}}, \bibnamefont{and}
  \bibinfo{author}{\bibfnamefont{U.}~\bibnamefont{Alon}}, \bibinfo{journal}{BMC
  evolutionary biology} \textbf{\bibinfo{volume}{7}}, \bibinfo{pages}{169}
  (\bibinfo{year}{2007}).

\bibitem[{\citenamefont{Kreimer et~al.}(2008)\citenamefont{Kreimer, Borenstein,
  Gophna, and Ruppin}}]{kreimer2008}
\bibinfo{author}{\bibfnamefont{A.}~\bibnamefont{Kreimer}},
  \bibinfo{author}{\bibfnamefont{E.}~\bibnamefont{Borenstein}},
  \bibinfo{author}{\bibfnamefont{U.}~\bibnamefont{Gophna}}, \bibnamefont{and}
  \bibinfo{author}{\bibfnamefont{E.}~\bibnamefont{Ruppin}},
  \bibinfo{journal}{Proc. Natl. Acad. Sci. U.S.A.}
  \textbf{\bibinfo{volume}{105}}, \bibinfo{pages}{6976} (\bibinfo{year}{2008}).

\bibitem[{\citenamefont{Singh et~al.}(2008)\citenamefont{Singh, wolf, Wang, and
  Arkin}}]{singh2008}
\bibinfo{author}{\bibfnamefont{A.~H.} \bibnamefont{Singh}},
  \bibinfo{author}{\bibfnamefont{D.~M.} \bibnamefont{wolf}},
  \bibinfo{author}{\bibfnamefont{P.}~\bibnamefont{Wang}}, \bibnamefont{and}
  \bibinfo{author}{\bibfnamefont{A.~P.} \bibnamefont{Arkin}},
  \bibinfo{journal}{PNAS} \textbf{\bibinfo{volume}{105}}, \bibinfo{pages}{7500}
  (\bibinfo{year}{2008}).

\bibitem[{\citenamefont{Sobolevsky and Trifonov}(2005)}]{r7}
\bibinfo{author}{\bibfnamefont{Y.}~\bibnamefont{Sobolevsky}} \bibnamefont{and}
  \bibinfo{author}{\bibfnamefont{E.~N.} \bibnamefont{Trifonov}},
  \bibinfo{journal}{J. Mol. Evol.} \textbf{\bibinfo{volume}{61}},
  \bibinfo{pages}{591} (\bibinfo{year}{2005}).

\bibitem[{\citenamefont{Hedges}(2002)}]{r1}
\bibinfo{author}{\bibfnamefont{S.~B.} \bibnamefont{Hedges}},
  \bibinfo{journal}{Nature} \textbf{\bibinfo{volume}{3}}, \bibinfo{pages}{838}
  (\bibinfo{year}{2002}).

\bibitem[{\citenamefont{Reid et~al.}(2000)\citenamefont{Reid, Herbelin,
  Bumbaugh, Selander, and Whittam}}]{Reid}
\bibinfo{author}{\bibfnamefont{S.~D.} \bibnamefont{Reid}},
  \bibinfo{author}{\bibfnamefont{C.~J.} \bibnamefont{Herbelin}},
  \bibinfo{author}{\bibfnamefont{A.~C.} \bibnamefont{Bumbaugh}},
  \bibinfo{author}{\bibfnamefont{R.~K.} \bibnamefont{Selander}},
  \bibnamefont{and} \bibinfo{author}{\bibfnamefont{T.~S.}
  \bibnamefont{Whittam}}, \bibinfo{journal}{Nature}
  \textbf{\bibinfo{volume}{406}}, \bibinfo{pages}{64} (\bibinfo{year}{2000}).

\bibitem[{\citenamefont{Hirsh et~al.}(2005)\citenamefont{Hirsh, Fraser, and
  Wall}}]{hirsh}
\bibinfo{author}{\bibfnamefont{A.~E.} \bibnamefont{Hirsh}},
  \bibinfo{author}{\bibfnamefont{H.~B.} \bibnamefont{Fraser}},
  \bibnamefont{and} \bibinfo{author}{\bibfnamefont{D.~P.} \bibnamefont{Wall}},
  \bibinfo{journal}{Molecular Biology and Evolution}
  \textbf{\bibinfo{volume}{22}}, \bibinfo{pages}{174} (\bibinfo{year}{2005}).

\bibitem[{\citenamefont{Apic et~al.}(2001)\citenamefont{Apic, Gough, and
  Teichmann}}]{r16}
\bibinfo{author}{\bibfnamefont{G.}~\bibnamefont{Apic}},
  \bibinfo{author}{\bibfnamefont{J.}~\bibnamefont{Gough}}, \bibnamefont{and}
  \bibinfo{author}{\bibfnamefont{S.~A.} \bibnamefont{Teichmann}},
  \bibinfo{journal}{J. Mol. Biol.} \textbf{\bibinfo{volume}{310}},
  \bibinfo{pages}{311} (\bibinfo{year}{2001}).

\bibitem[{\citenamefont{Ng et~al.}(2003{\natexlab{a}})\citenamefont{Ng, Zhang,
  Tan, and Lin}}]{r23}
\bibinfo{author}{\bibfnamefont{S.}~\bibnamefont{Ng}},
  \bibinfo{author}{\bibfnamefont{Z.}~\bibnamefont{Zhang}},
  \bibinfo{author}{\bibfnamefont{S.}~\bibnamefont{Tan}}, \bibnamefont{and}
  \bibinfo{author}{\bibfnamefont{K.}~\bibnamefont{Lin}},
  \bibinfo{journal}{Nucl. Acids Res.} \textbf{\bibinfo{volume}{31}},
  \bibinfo{pages}{251} (\bibinfo{year}{2003}{\natexlab{a}}).

\bibitem[{\citenamefont{Ng et~al.}(2003{\natexlab{b}})\citenamefont{Ng, Zhang,
  and Tan}}]{r24}
\bibinfo{author}{\bibfnamefont{S.}~\bibnamefont{Ng}},
  \bibinfo{author}{\bibfnamefont{Z.}~\bibnamefont{Zhang}}, \bibnamefont{and}
  \bibinfo{author}{\bibfnamefont{S.}~\bibnamefont{Tan}},
  \bibinfo{journal}{Bioinformatics} \textbf{\bibinfo{volume}{19}},
  \bibinfo{pages}{923} (\bibinfo{year}{2003}{\natexlab{b}}).

\bibitem[{\citenamefont{Ravasz et~al.}(2002)\citenamefont{Ravasz, Somera,
  Mongru, Oltvai, and Barab\'{a}si}}]{r26}
\bibinfo{author}{\bibfnamefont{E.}~\bibnamefont{Ravasz}},
  \bibinfo{author}{\bibfnamefont{A.~L.} \bibnamefont{Somera}},
  \bibinfo{author}{\bibfnamefont{D.~A.} \bibnamefont{Mongru}},
  \bibinfo{author}{\bibfnamefont{Z.~N.} \bibnamefont{Oltvai}},
  \bibnamefont{and} \bibinfo{author}{\bibfnamefont{A.~L.}
  \bibnamefont{Barab\'{a}si}}, \bibinfo{journal}{Science}
  \textbf{\bibinfo{volume}{297}}, \bibinfo{pages}{1551} (\bibinfo{year}{2002}).

\bibitem[{\citenamefont{Eisen et~al.}(1998)\citenamefont{Eisen, Spellman,
  Brown, and Botstein}}]{Botstein}
\bibinfo{author}{\bibfnamefont{M.~B.} \bibnamefont{Eisen}},
  \bibinfo{author}{\bibfnamefont{P.~T.} \bibnamefont{Spellman}},
  \bibinfo{author}{\bibfnamefont{P.~O.} \bibnamefont{Brown}}, \bibnamefont{and}
  \bibinfo{author}{\bibfnamefont{D.}~\bibnamefont{Botstein}},
  \bibinfo{journal}{Proc. Natl. Acad. Sci. USA} \textbf{\bibinfo{volume}{95}},
  \bibinfo{pages}{14863} (\bibinfo{year}{1998}).

\bibitem[{\citenamefont{Shapiro}(2002)}]{Shapiro3}
\bibinfo{author}{\bibfnamefont{J.~A.} \bibnamefont{Shapiro}},
  \bibinfo{journal}{J. Biol. Phys.} \textbf{\bibinfo{volume}{28}},
  \bibinfo{pages}{745} (\bibinfo{year}{2002}).

\bibitem[{\citenamefont{Colegrave}(2002)}]{Colegrave}
\bibinfo{author}{\bibfnamefont{N.}~\bibnamefont{Colegrave}},
  \bibinfo{journal}{Nature} \textbf{\bibinfo{volume}{420}},
  \bibinfo{pages}{664} (\bibinfo{year}{2002}).

\bibitem[{\citenamefont{Goddard et~al.}(2005)\citenamefont{Goddard, Godfray,
  and Burt}}]{Burt}
\bibinfo{author}{\bibfnamefont{M.~R.} \bibnamefont{Goddard}},
  \bibinfo{author}{\bibfnamefont{H.~C.~J.} \bibnamefont{Godfray}},
  \bibnamefont{and} \bibinfo{author}{\bibfnamefont{A.}~\bibnamefont{Burt}},
  \bibinfo{journal}{Nature} \textbf{\bibinfo{volume}{434}},
  \bibinfo{pages}{636} (\bibinfo{year}{2005}).

\bibitem[{\citenamefont{Soyer and Bonhoeffer}(2006)}]{Bonhoeffer}
\bibinfo{author}{\bibfnamefont{O.~S.} \bibnamefont{Soyer}} \bibnamefont{and}
  \bibinfo{author}{\bibfnamefont{S.}~\bibnamefont{Bonhoeffer}},
  \bibinfo{journal}{Proc. Natl. Acad. Sci. USA} \textbf{\bibinfo{volume}{103}},
  \bibinfo{pages}{16337} (\bibinfo{year}{2006}).

\bibitem[{\citenamefont{Walsh}(2004)}]{Walsh}
\bibinfo{author}{\bibfnamefont{C.~T.} \bibnamefont{Walsh}},
  \bibinfo{journal}{Science} \textbf{\bibinfo{volume}{303}},
  \bibinfo{pages}{1805} (\bibinfo{year}{2004}).

\bibitem[{\citenamefont{Maiden}(1998)}]{Maiden}
\bibinfo{author}{\bibfnamefont{M.~C.~J.} \bibnamefont{Maiden}},
  \bibinfo{journal}{Clin. Inf. Dis.} \textbf{\bibinfo{volume}{27S}},
  \bibinfo{pages}{S12} (\bibinfo{year}{1998}).

\end{thebibliography}

\end{document}